\documentclass[twocolumn]{aastex63}

\usepackage{amsmath}

\accepted{\today}
\shorttitle{CSP-I: Host galaxies}
\shortauthors{Uddin et al.}
\usepackage{parskip}

\usepackage{hyperref}

\usepackage{longtable}
\usepackage{tabularx}
\usepackage{enumitem}

\begin{document}

\title{The Carnegie Supernova Project-I: Correlation Between Type Ia Supernovae and Their Host Galaxies from Optical to Near-Infrared Bands\footnote{This paper includes data gathered with the 6.5 meter Magellan Telescopes located at Las Campanas Observatory, Chile.}}

\correspondingauthor{Syed~A.~Uddin}
\email{suddin@carnegiescience.edu}

\author[0000-0002-0786-7307]{Syed~A~Uddin}
\affiliation{Observatories of the Carnegie Institution for Science, 813 Santa Barbara St, Pasadena, CA, 91101, USA}
\affiliation{School of Science, Technology, Engineering and Math, American Public University System, 111 W. Congress Street, Charles Town, WV 25414, USA }

%\collaboration{1}{(Carnegie Supernova Project Collaboration)}

\author[0000-0003-4625-6629]{Christopher~R.~Burns}
\affiliation{Observatories of the Carnegie Institution for Science, 813 Santa Barbara St, Pasadena, CA, 91101, USA}

\author[0000-0003-2734-0796]{M.~M.~Phillips}
\affiliation{Carnegie Observatories, Las Campanas Observatory, Casilla 601, La Serena, Chile}

\author[0000-0002-8102-181X]{Nicholas~B.~Suntzeff}
\affiliation{George P. and Cynthia Woods Mitchell Institute for Fundamental Physics and Astronomy, Texas A\&M University, College Station, TX, 77843, USA}
\affiliation{Department of Physics and Astronomy, Texas A\&M University, College Station, TX, 77843, USA}

\author[0000-0001-6293-9062]{Carlos~Contreras}
\affiliation{Carnegie Observatories, Las Campanas Observatory, Casilla 601, La Serena, Chile}

\author[0000-0003-1039-2928]{Eric~Y.~Hsiao}
\affiliation{Department of Physics, Florida State University, Tallahassee, FL 32306, USA}

\author[0000-0003-2535-3091]{Nidia~Morrell}
\affiliation{Carnegie Observatories, Las Campanas Observatory, Casilla 601, La Serena, Chile}

\author[0000-0002-1296-6887]{Llu\'is Galbany}
\affiliation{Departamento de F\'isica Te\'orica y del Cosmos, Universidad de Granada, E-18071 Granada, Spain}

\author[0000-0002-5571-1833]{Maximilian~Stritzinger}
\affiliation{Department of Physics and Astronomy, Aarhus University, Ny Munkegade 120, DK-8000 Aarhus C, Denmark}

\author[0000-0002-4338-6586]{Peter~Hoeflich}
\affiliation{Department of Physics, Florida State University, Tallahassee, FL 32306, USA}

\author[0000-0002-5221-7557]{Chris~Ashall}
\affiliation{Department of Physics, Florida State University, Tallahassee, FL 32306, USA}

\author[0000-0001-6806-0673]{Anthony~L.~Piro}
\affiliation{Observatories of the Carnegie Institution for Science, 813 Santa Barbara St, Pasadena, CA, 91101, USA}

\author[0000-0003-3431-9135]{Wendy~L.~Freedman}
\affiliation{Department of Astronomy \& Astrophysics, University of Chicago, 5640 South Ellis Avenue, Chicago, IL 60637, USA}

\author[0000-0003-0554-7083]{S.~E.~Persson}
\affiliation{Observatories of the Carnegie Institution for Science, 813 Santa Barbara St, Pasadena, CA, 91101, USA}

%\author{Mario Hamuy}
%\affiliation{Universidad de Chile, Departmento de Astronomia, Casilla 36-D, Santiago, Chile}

\author[0000-0002-6650-694X]{Kevin~Krisciunas}
\affiliation{George P. and Cynthia Woods Mitchell Institute for Fundamental Physics and Astronomy, Texas A\&M University, College Station, TX, 77843, USA}

\author[0000-0001-6272-5507]{Peter~Brown}
\affiliation{George P. and Cynthia Woods Mitchell Institute for Fundamental Physics and Astronomy, Texas A\&M University, College Station, TX, 77843, USA}



\begin{abstract}

%\textcolor{red}{FINAL FINAL FINAL FINAL VERSION}\\

We present optical and near-infrared ($ugriYJH$) photometry of host galaxies of
Type Ia supernovae (SN~Ia) observed by the  \textit{Carnegie Supernova
Project-I}. We determine host galaxy stellar masses and, for the first time,
study their correlation with SN~Ia standardized luminosity across optical
and near-infrared ($uBgVriYJH$) bands. In the individual bands, we find that SNe~Ia are more luminous in more massive hosts with luminosity offsets ranging between $-0.07 \pm0.03$ mag to $-0.15\pm0.04$ mag after light-curve standardization. The slope
of the SN~Ia Hubble residual-host mass relation is negative across all
$uBgVriYJH$ bands with values ranging between $-0.029\pm 0.029$ mag/dex to $-0.093\pm 0.031$ mag/dex -- implying that SNe~Ia in more massive galaxies are
brighter than expected. 
The near-constant observed correlations across optical and near-infrared bands indicate that
dust may not play a significant role in the observed
luminosity offset--host mass correlation.  We measure projected separations
between SNe~Ia and their host centers, and find that SNe~Ia that explode
beyond a projected 10 kpc 
have a $\rm 50\% \ to \ 60\%$ reduction of
the dispersion in Hubble residuals across all bands --- making them a more uniform subset of SNe~Ia.
Dust in host galaxies, peculiar velocities of nearby SN~Ia, or a combination of both may drive this result as the color excesses of
SNe~Ia beyond 10 kpc are found to be generally lower
than those interior, but there is also a diminishing trend of the dispersion as we exclude nearby events.  
We do not find that SN~Ia average luminosity varies significantly when they are grouped in
various host morphological types. Host galaxy data from this work will be
useful, in conjunction with future high-redshift samples, in constraining
cosmological parameters.  

%Combining our data with the Sloan Digital Sky Survey supernova program and the Supernova Legacy Survey improves statistical significance of this correlation, resulting in a luminosity offset of $ 0.07  \ (\sim  2.1\sigma)$ mag, and a negative slope of $-0.065 \ (6\sigma)$ mag/dex.

\end{abstract}

%\keywords{supernovae: general, galaxies: fundamental parameters,
%galaxies: photometry, cosmology:distance scale}


\section{Introduction} \label{sec:intro}

Since the surprising discovery of the accelerated expansion of the Universe
(\citealt{riess98}, \citealt{perlmutter97}) using the standardization of
SNe~Ia (\citealt{phillips93}, \citealt{tripp98}), we have made little progress
in understanding the physical cause of this observed acceleration. Although
recent results tend to point towards a value consistent with the cosmological
constant, other explanations are also possible (see \citealt{joyce16}).
Alternative possibilities such as the time-varying nature of dark energy are
currently difficult to test due to systematic uncertainties in deriving
cosmological parameters (e.g., \citealt{betoule14}, \citealt{scolnic17}). Along
with improving photometric calibration, systematics of astrophysical origin
need more attention. One such systematic is the interplay between the
properties of SNe~Ia and their hosts. 

A number of studies have been made on the correlation between the properties of SNe~Ia and their
hosts (\citealt{hamuy95}, \citealt{sullivan10}, \citealt{lampeitl10a},
\citealt{childress13}, \citealt{pan14}, \citealt{uddin17b}, \citealt{roman18},
\citealt{wiseman20}). These studies can be grouped into two categories: global
host properties and local host properties (see \citealt{uddin17b} for a
review). All studies, with varying degrees of significance, show that 
on average SNe~Ia
%peak brightness 
decline faster and are relatively over-luminous (i.e.,
more negative values of Hubble residuals\footnote{Hubble residual is the difference
% CRB: To make more luminous SNe negative, it has to be the difference
% between the predicted and observed
between the predicted and the observed values of distance moduli after
obtaining a best-fit cosmological model for a given set of SNe~Ia. We will
refer to Hubble residuals as luminosities in such a way that negative values 
%of Hubble residuals means 
indicate more luminous SNe~Ia.}) after light-curve
standardization when hosted in massive galaxies. A few studies show
correlations between SN~Ia properties and SN~Ia--host galaxy separation, as well as
morphological types (\citealt{wang97}, \citealt{hicken09}, and \citealt{galbany12}). Some of these studies find that certain subgroups of
SNe~Ia can be better standard candles based on their host properties (e.g.,
\citealt{rigault13}, \citealt{kelly15}, \citealt{uddin17b}). Moreover,
\cite{uddin17a} confirmed there is no bias in cosmological constraints derived from
subsamples selected by host properties.  

%\citet{wang13} found that, SNe~Ia with high-velocity ejecta tend to explode near the inner and brighter region of their host galaxies.

The correlation between SNe~Ia luminosities and their hosts currently lacks explanation.
%The existence of a correlation between SNe~Ia and their hosts needs an
%explanation.
 A possible suspect is the 
%role of dust in driving the
%luminosity-host mass correlation 
effect of dust in the host galaxies \citep{bs20}. One way to test 
%the role 
%of dust
this is to study SN~Ia luminosity--host stellar mass correlations in multiple bands,
preferably from optical to near-infrared. All previous studies, except
\citet{burns18}\footnote{We note that \citet{burns18} use 2MASS $K$-band photometry to determine host galaxy stellar mass: details in \autoref{sec:prop}.}, have used $B$-band peak brightness to study SN~Ia--host correlations. 
In this study, we determine $ugriYJH$ photometry and derive physical properties of host galaxies of SNe~Ia from
the \textit{Carnegie Supernova Project-I} (hereafter CSP-I;
\citealt{hamuy06}). Thereafter, we study the correlation of SN~Ia Hubble residuals 
in $uBgVriYJH$ bands with
these properties. This paper is the first to study SN~Ia Hubble
residual-host stellar mass correlations from optical to near-infrared
wavelengths.

%Somewhere in the intro, you should point out that most host-mass/Luminosity correlations are done for a single band (B,V, ie. optical), but I don’t think anyone has investigated the correlation as a function of wavelength. The motivation is that it could be a reddening thing (maybe dust is systematically different in low vs. high-mass hosts). This gives the reader assurance that this isn’t just the same old thing. 

The CSP-I was designed to follow SNe~Ia discovered at low redshifts, mostly from targeted
transient search programs. The CSP-I performed photometric follow-ups over a wide
range of bands from optical to near-infrared ($uBgVriYJH$), and was conducted
between 2004-2010. Light-curves of SNe~Ia from CSP-I are published in three
papers (\citealt{contreras10}, \citealt{stritzinger11}, and
\citealt{krisciunas17}). \cite{burns18} presented an absolute calibration of CSP-I
SNe~Ia photometry, and a new measurement of the Hubble constant. The second
phase of the CSP (CSP-II) was conducted between 2011-2015, and the resulting SN~Ia
sample characteristics are described in \citet{phillips18} and in
\citet{hsiao19}. Optical and near-infrared Hubble diagrams from a combined
SN~Ia sample from CSP-I and CSP-II will be presented in a future paper (Uddin
et al. in preparation). The purpose of this paper is to present the methodology
of host mass determination and apply this to publicly available CSP-I data.

In \autoref{sec:data}, we describe the SN~Ia and host galaxy sample, 
\autoref{sec:prop}
describes host galaxy properties, and \autoref{sec:corr} presents correlations
between SNe~Ia and hosts. A discussion of our findings is in \autoref{sec:dis},
and we summarize our conclusions in \autoref{sec:con}.

\section{Data} \label{sec:data}

\subsection{SNe~Ia Light-curves and Distances}\label{sec:lc}

We fit SN~Ia light-curves using SNooPy (\citealt{burns11}). Briefly, the
best-observed SNe~Ia from the CSP-I are used as a training set to obtain
light-curve templates for the set of $uBgVriYJH$ filters as a function of a
dimensionless color-stretch parameter $s_{BV}$, which is defined as
the rest-frame time of maximum of the $B-V$ color-curve relative to time of
$B$-band maximum divided by 30 \citep{burns14}. These templates are then
used to fit the CSP-I sample of SNe~Ia, resulting in estimates of $s_{BV}$, and
magnitudes at maximum in each filter. Using the intrinsic colors 
from \citet{burns14}, the reddening parameters $E(B-V)_{\rm host}$ and $R_V^{\rm host}$ 
can also be determined. Finally, using the absolute magnitudes from
\citet{burns18}, we can infer distances and compute Hubble residuals. We describe details below.

Since we wish to investigate the behavior of Hubble residuals as a function of wavelength, it is important to minimize, as much as possible, the correlation between different filters. To this end, we fit each SN with SNooPy's \texttt{max$\_$model}. This model uses the light-curve shape templates as a function of color stretch $s_{BV}$ to fit the maximum brightness $m_{max,\lambda}$  in each filter {\em independently}. However, the magnitudes are corrected for Milky~Way extinction using the \citet{sf11} dust maps, which introduces small correlated errors that are unavoidable. SNooPy also applies k-corrections computed using the \citet{hsiao07} SN~Ia spectral sequence.

With these magnitudes at maximum in $N$ filters, we can construct $N-1$ independent colors at maximum. Using the methods from \citet{burns14}, we use these colors to infer best-fit values for the host galaxy color excess $E(B-V)_{\rm host}$ and reddening slope $R_V^{\rm host}$  for each SN. Again, this introduces some correlated errors due to the uncertainties in the extinction parameters which will be largest in the optical and smaller in the NIR. It is important that only colors are used to constrain the extinction. As shown in \citet{burns18}, determining extinction by minimizing Hubble residuals leads to a significant bias in $E(B-V)_{\rm host}$ and $R_V^{\rm host}$.

Once the apparent magnitudes at maximum have been corrected for extinction, we compute absolute magnitudes using the luminosity distance\footnote{ Luminosity distances are computed assuming a Hubble--Lema$\rm \hat{i}$tre constant $H_0 = 72 \mathrm {km\ s^{-1}\ Mpc^{-1}}$ and deceleration parameter $q_0 = -0.53$. For more details, see section 5.1 of \citet{burns18}.}, $\mu_L$ (or Cepheid distance if available). We then fit these absolute magnitudes as a function of $s_{BV}$ to obtain a luminosity-decline rate relation for each filter independently, using a 3rd-order polynomial (see \citealt{burns18}). To measure Hubble residuals, we use leave-one-out cross-validation: for each of the $M$ SNe~Ia, we use the other $M-1$ to solve for the luminosity-decline rate relation. We use this to infer the distance to the SN that was left out, $\mu_{\rm SN}$. This inferred distance is subtracted from the $\mu_L$ to obtain the Hubble residual, $\mu_{\rm SN} - \mu_L$ and we express it as $\Delta \mu$. The procedure is repeated for each filter and for all SNe.

Despite all these measures, there remains a large degree of correlation between the residuals of different filters. While most of this is due to the correlated brightness of the SNe themselves, at the median redshift of 0.025 for the CSP-I, an average peculiar velocity of 300 km/s would introduce an additional correlation of +/- 0.1 mag and could get as high as +/- 0.5 mag for our lowest-redshift objects. The latter can be mitigated by considering different redshift cuts.
%The majority of this correlation is due to the effects of peculiar velocities, which introduce achromatic errors in distance. 
We attempt to mitigate this by considering different redshift cuts. We also attempt to remove the fastest decliners (for which the  luminosity-decline rate relation corrections are largest) and the reddest objects (to reduce correlations due to extinction corrections). As we make more and more of these cuts, however, the statistical errors increase and we lose the ability to say anything definitive about the wavelength dependence of these mass effects (see \autoref{hr}). 

While the correlations due to peculiar velocity could be a cause for concern, we have no reason to believe they would be biased with respect to the mass of the host galaxy. They therefore introduce a significant level of white noise, but should not introduce a mass step/slope bias. Nevertheless, a more definitive answer will be possible using a sample like CSP-II that has a higher mean redshift (\citealt{phillips18}), is not as biased to high-mass galaxy hosts, and has sufficient numbers to allow selection cuts in decline rate and/or extinction.

The Hubble
residuals of the CSP-I SN~Ia sample used in this work are tabulated in Table
\ref{restable}. The following peculiar SNe Ia from the CSP-I are excluded: SN~2004dt,
SN~2005gj, SN~2005hk, SN~2006bt, SN~2006ot,  SN~2007so, SN~2008ae, SN~2008bd,
SN~2008ha, SN~2008J, SN~2009dc, SN~2009J, and SN~2010ae \citep[for details, see][]{krisciunas17}.

\subsection{Host Galaxies}\label{sec:obs}

We obtain the list of host galaxies of CSP-I SNe~Ia from \citet{krisciunas17}.
This is the final data release paper of CSP-I photometry and lists 123 SNe~Ia in the
redshift range of $0.004 < z < 0.083$. Host galaxy reference images were obtained at
sufficiently later times so that no significant contribution of light from the
SNe~Ia contaminates the photometry. These were
obtained primarily using the Tek5 optical CCD camera and the near-infrared WIRC (and RetroCam from 2011) on the 2.5~m
Ir\`{e}n\`{e}e du Pont telescope, as well as with the SITe3 CCD camera and the near-infrared RetroCam on the 1~m Henrietta Swope
telescopes at the Las Campanas Observatory. Some near-infrared
images were also taken with the FourStar instrument on the 6.5~m Magellan Baade
telescope. The reader is referred to \citet{krisciunas17} for further details.

In several cases, host identification of the SN~Ia remains ambiguous.
As described in detail in \citet{krisciunas17}, the host galaxies of SN~2004gc,
SN~2007A, SN~2007if, SN~2007mm, SN~2008bf, and SN~2008ff are unclear.
These SNe~Ia exploded within groups of galaxies where it is difficult to
assign the true host, and so we exclude them from this paper. Also, two SNe~Ia
share the same host: SN~2006mr and SN~2006dd. We exclude SN~2006dd since it was not 
observed by the CSP-I until very late epochs, making light curve fitting with SNooPy impossible.
Finally, we exclude SN~2007sr as the host galaxy of this SN~Ia is NGC 4038/NGC4039, also known as the Antennae Galaxies.
SN~2007sr exploded in one of the tidal tails of this interactive pair
(\citealt{schweizer08}), which makes it difficult to assign the correct host.
%Moreover, the field of view of the CCD camera of the du Pont Telescope was not large enough to image this pair of galaxies. 

%Table~\ref{hostlist} summarizes selection of hosts that we use in this paper. 
%
%\begin{table}[htp]
%\caption{CSP-I Host Galaxy Sample used in this work.}
%\begin{center}
%\begin{tabular}{lc}
%\hline
%\hline
%Criteria & Sample\\
%\hline
%Initial (\citealt{krisciunas17}) & 134\\
%Six ambiguous hosts & 6\\
%Single host for SN~2006mr and SN~2006dd & 1\\
%Excluding host of SN~2007sr & 1\\
%\hline
%Total for this work & 126\\
%\hline
%\end{tabular}
%\end{center}
%\label{hostlist}
%\end{table}%
%

%\begin{figure}[htbp]
%\begin{center}
%\includegraphics[width=\columnwidth]{ExpofImage.pdf}
%\caption{Distributions of exposures of host galaxies in CSPI.}
%\label{exp}
%\end{center}
%\end{figure}
%
%\begin{figure}[htbp]
%\begin{center}
%\includegraphics[width=\columnwidth]{NoofImage.pdf}
%\caption{Distributions of number of images went in making deep coadded images in CSPI.}
%\label{noim}
%\end{center}
%\end{figure}
%

\subsection{Host Galaxy Photometry}\label{sec:phot}
We measure apparent magnitudes of the host galaxies in each of the $ugriYJH$ bands. Before
performing photometry, we align all images with respect to the centers of
each host galaxy, then co-add multiple exposures to make deeper images.
We perform photometry using Source Extractor (SExtractor; \citealt{bertin96})
in single image mode. We inspect SExtractor check images for failed detection
and appropriate deblending. Using calibrated local sequence stars, we determine
zero points in each stacked image and compute apparent magnitudes of host
galaxies using the 
\texttt{MAG$\_$AUTO} magnitude definition of SExtractor. 

\begin{figure}[htbp]
\begin{center}
\includegraphics[width=\columnwidth]{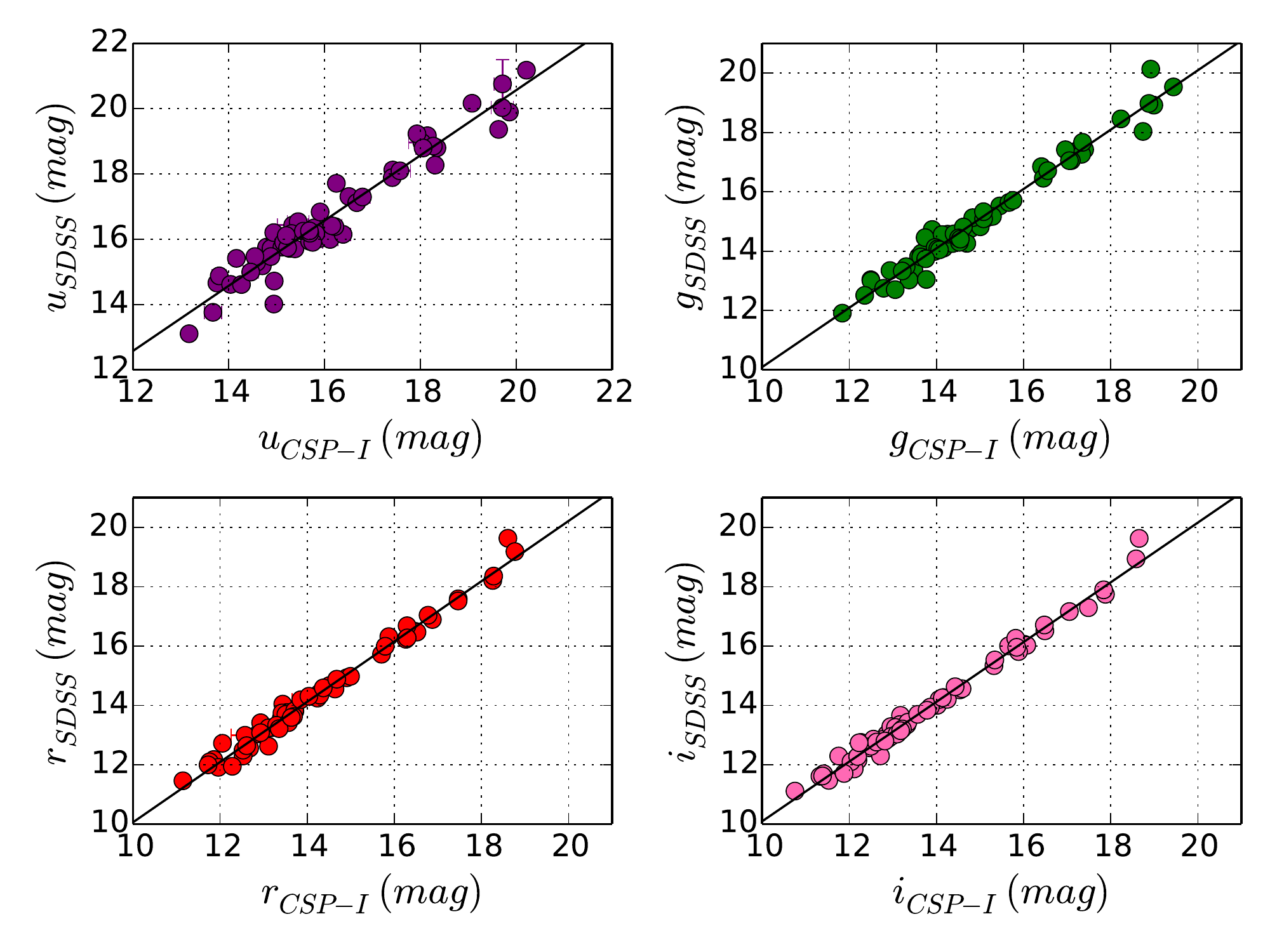} 
\caption{Comparison between host galaxy magnitudes from CSP and SDSS. In each
   case, solid lines are the best-fit linear relations to the data.}
\label{sdss} \end{center} \end{figure}

\begin{figure*}[htbp]
\begin{center}
\begin{tabular}{cc}

\includegraphics[width=\columnwidth]{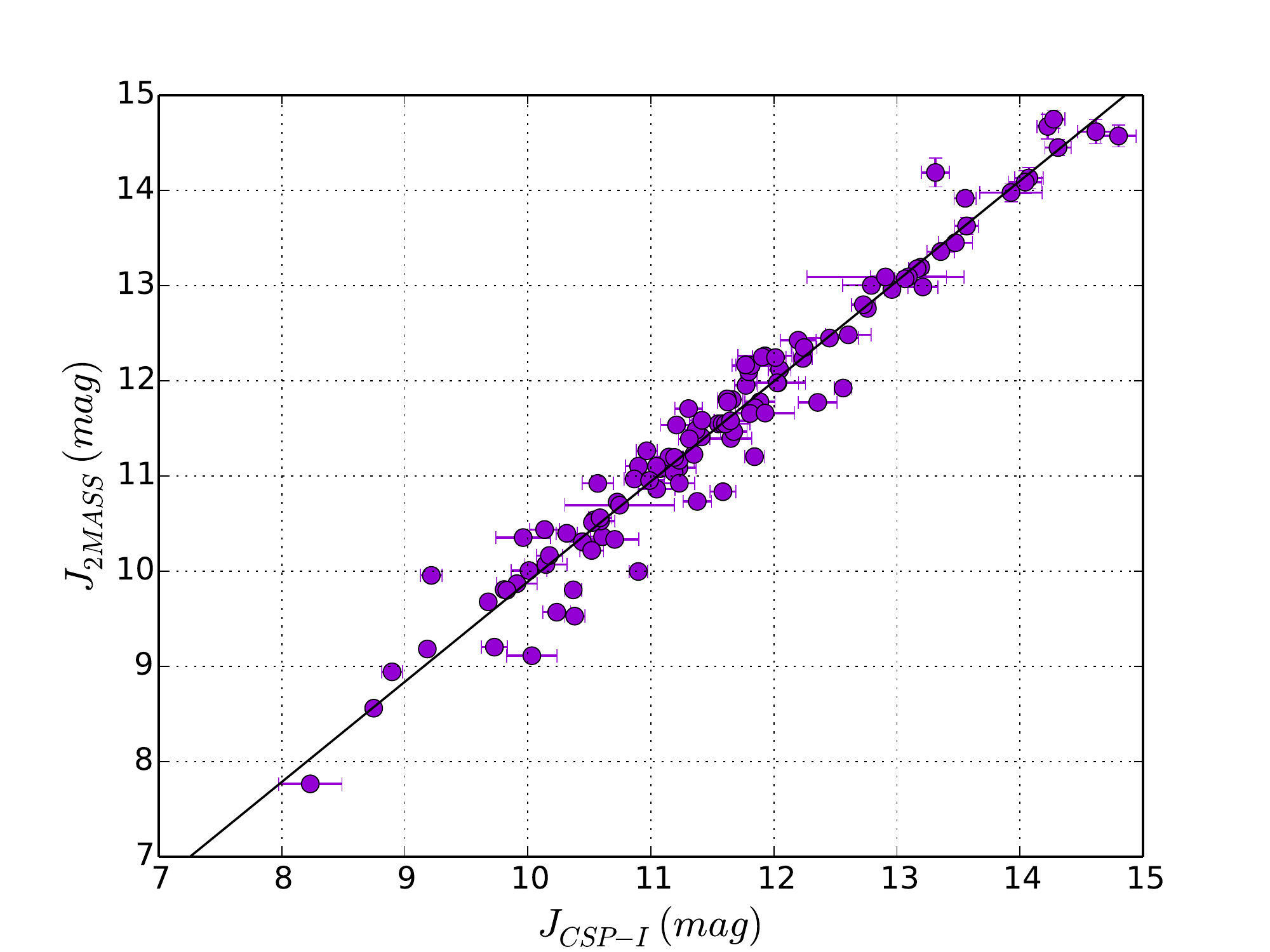} &
\includegraphics[width=\columnwidth]{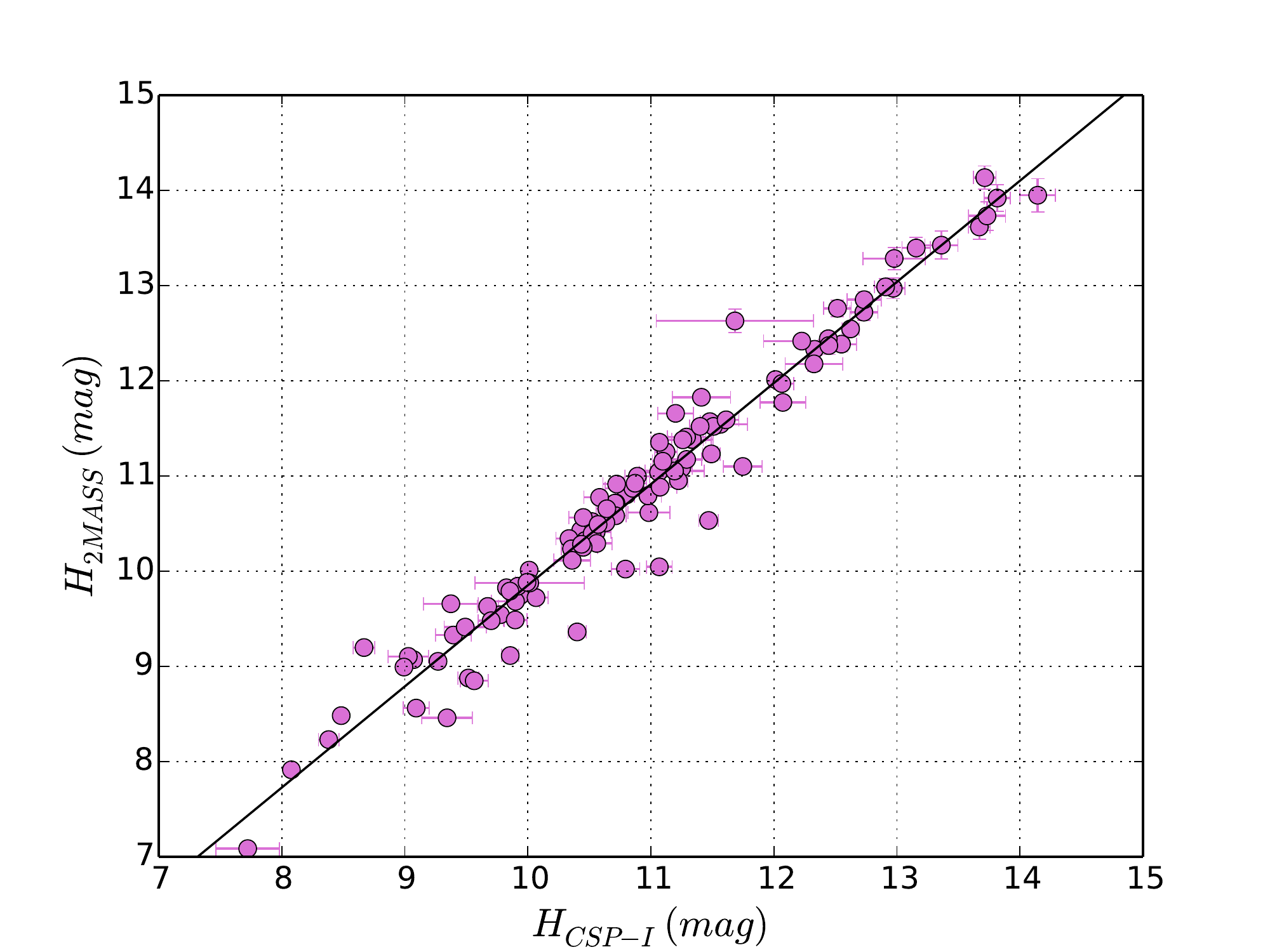}
\end{tabular}

\caption{Comparison between host galaxy magnitudes from CSP-I and the 2MASS
extended source catalog. In each case, solid lines are the best-fit linear
   relations to the data (see Eqn.~2).
} \label{2mass} \end{center} \end{figure*}

As a check, we compare our optical photometry with the SDSS model magnitudes 
available from the
SDSS Sky
Server\footnote{https://skyserver.sdss.org/dr12/en/tools/search/IQS.aspx}, and with
near-infrared photometry ($J$ and $H$) from the 2MASS all-sky survey extended
source catalog\footnote{http://tdc-www.harvard.edu/catalogs/tmxsc.html}. We
show these comparisons in Fig.~\ref{sdss} and in Fig.~\ref{2mass},
respectively. Slopes of best-fit relations and mean offsets with respect to CSP-I photometry are shown in Table~\ref{off}.

\begin{table}[htp]
\caption{Slopes of best-fit relations and mean offsets when comparing CSP-I  photometry with SDSS $ugri$ photometry, and  with 2MASS $JH$ photometry. }
\begin{center}
\begin{tabular}{|c|c|c|}
\hline
Band & Slope & Offset (mag)\\
\hline
$u$ & 0.99&0.18\\
$g$ & 1.00&0.05\\
$r$ & 1.02&0.06\\
$i$ & 1.01&0.06\\
$J$ & 1.04&0.03\\
$H$ & 1.04&0.03\\
\hline
\end{tabular}
\end{center}
\label{off}
\end{table}%

%Best-fit relations between SDSS and CSP-I are:
%
%\scriptsize
%\begin{equation}
%  \textit{SDSS = CSP-I}
%    \begin{cases}
%      \text{$m=0.99; c=0.31$}, &   \text{($u-15.68$)}\\
%      \text{$m=1.00; c=0.08$}, & \text{($g$)}\\
%      \text{$m=1.02; c=0.10$},& \text{($r$)}\\
%      \text{$m=1.01; c=0.02$}. &  \text{($i$)}
%    \end{cases}       
%\end{equation}
%\normalsize
%
%where $m$ is the slope and $c$ is the mean offset between two photometric systems. Between 2MASS and CSP-I we obtain:
%\scriptsize
%\begin{equation}
%  \textit{2MASS = CSP-I}
%    \begin{cases}
%      \text{$m=1.04; c=-0.55$}, &   \text{($J$)}\\
%      \text{$m=1.04; c=-0.49$}. & \text{($H$)}\\
%         \end{cases}       
%\end{equation}
%\normalsize
%
%The RMS dispersion in the above fits is $\sim 0.2$ mag across all bands. 
In some
cases, when the host galaxy was not observed in a particular band by the CSP, 
photometry
was obtained from external sources. For optical ($ugri$) filters, SDSS model
magnitudes\footnote{SDSS recommends the use of model magnitudes for extended
objects and when color information, useful for stellar mass estimation, is preferred; details are in https://www.sdss.org/dr12/algorithms/magnitudes/} were
used, and
for near-infrared ($J$ and $H$), 2MASS extended source magnitudes were used.
%Eqns.~1 and 2 
Offsets from Table~\ref{off} were applied to convert SDSS and 2MASS magnitudes to CSP-I system.
We present photometry of the CSP-I SN~Ia host galaxies in Table~\ref{phottable}.

%Fig.~\ref{mag} shows distributions of host galaxy absolute magnitudes.
%
%\begin{figure*}[htbp]
%\begin{center}
%\includegraphics[width=\textwidth]{AbsMagCSP1.pdf}
%\caption{Distribution of host galaxy absolute magnitudes of SNe~Ia from CSP-I in $ugriYJH$ bands.  }
%\label{mag}
%\end{center}
%\end{figure*}

% comapring with SDSS

% comparing with 2MASS 

\section{Host Galaxy Properties}\label{sec:prop}

\subsection{Stellar Mass}
\begin{figure}[htbp]
\begin{center}
%\begin{tabular}{cc}
\includegraphics[width=\columnwidth]{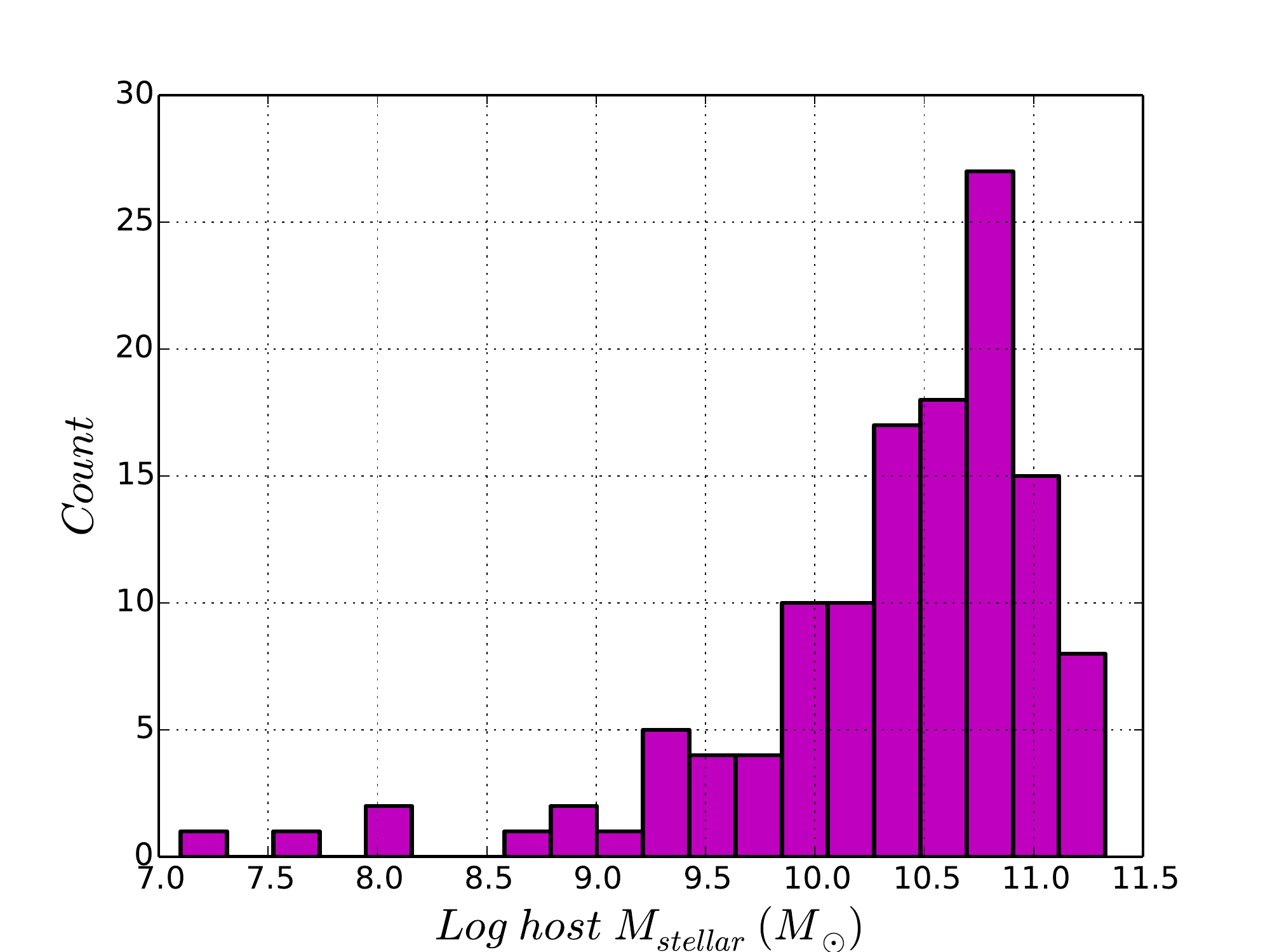} 
%\includegraphics[width=\columnwidth]{CSPI_sfrmass.pdf} 
%\end{tabular}
\caption{Distribution of host galaxy stellar mass of SNe~Ia from CSP-I. The
distribution has a peak at a relatively higher mass, which is a consequence of
galaxy-targeted SN~Ia searches.} \label{mass} \end{center} \end{figure}

The physical properties of the host galaxies are derived by using the Z-PEG
(\citealt{leborgne02}) Spectral Energy Distribution (SED) software to
fit the $ugriYJH$ photometry. We follow the same procedure as described in
\citet{uddin17b}.  Z-PEG uses the PEGASE 2.0 (\citealt{fioc97}) spectral
libraries, and computes best matches using $\chi^2$ minimization. There are
nine galaxy templates: E, S0, Sa, Sb, Sbc, Sc, Sd, Im, and starburst.
Additional internal extinction values are applied to each template ranging
from 0.0 mag to 0.5 mag in steps of 0.05 mag. The \citet{rana92} 
Initial Mass
Function was adopted. The stellar masses of the CSP-I host galaxies are shown in 
Fig.~\ref{mass} and
presented in Table~\ref{masstable}. The distribution has a peak at a 
relatively high
mass, which is consistent with the fact that most SNe~Ia were discovered by
galaxy-targeted supernova searches (\citealt{hamuy06}). In contrast, host galaxies of SNe~Ia from unbiased surveys have more low-mass galaxies (e.g., see figure 6 of \citealt{uddin17b}).

\cite{kelly10}  demonstrated that galaxy stellar masses determined from
multi-band photometry are in good agreement with those derived using galaxy
spectra. They  considered 10,000 galaxies from the SDSS and found an rms
dispersion of 0.15 dex when they compared stellar masses derived using $ugriz$
photometry to those using spectral indices. \cite{childress13a} also performed a
similar analysis and found similar dispersions when comparing 3673 galaxies.

\begin{figure}[htbp]
\begin{center}
\includegraphics[width=\columnwidth]{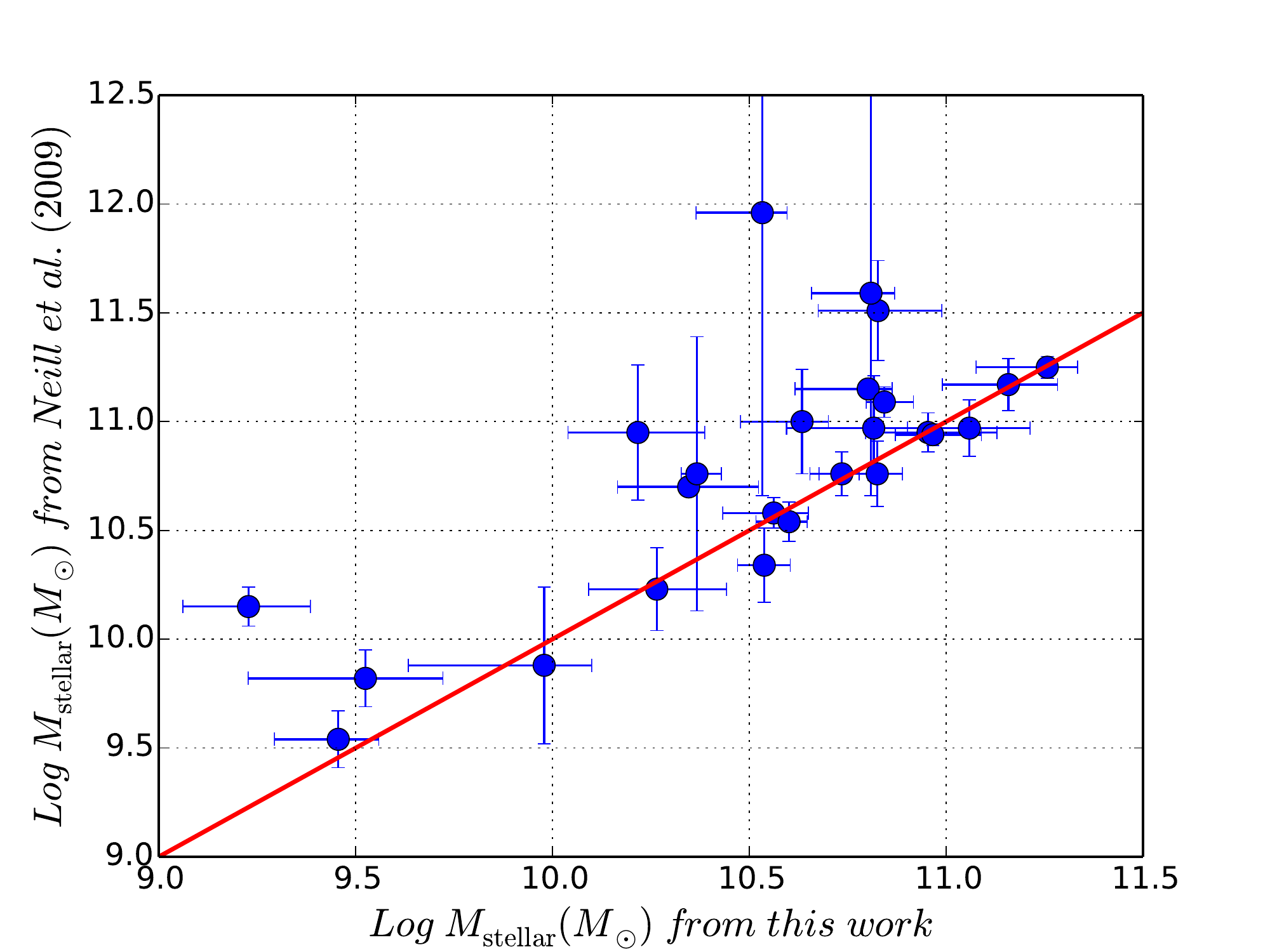}
\caption{Comparison of host stellar masses from CSP-I and \citet{neill09}. The
   solid line shows
   the one-to-one relation. \citet{neill09} determined, on average, 
   $0.18$ dex higher
   mass than what we calculate in this work. See text for discussion.}
   \label{neill} \end{center} \end{figure}

Previously, \citet{burns18} calculated host galaxy stellar masses for the
CSP-I SNe~Ia
using 2MASS $K$-band magnitudes. They utilized an empirical relation assuming
a constant stellar mass-to-light ratio and normalized to the stellar masses from
\cite{neill09}. The host stellar masses that we calculate from this work are on
average 0.3 dex lower than \citet{burns18}. Comparing the host masses
of 25 SNe~Ia from this work that are in common with \cite{neill09}, we find a
mean difference of 0.18 dex. This comparison is shown in Fig.~\ref{neill}.
It should  be noted that \cite{neill09} used additional ultra-violet photometry from
GALEX along with optical photometry in their analysis. %As discussed in 
%\cite{childress13a},ultra-violet photometry can be influenced by star-formation histories, and can lead
%to differences in deriving photometry-based galaxy properties.  
Only $40\%$ of the host galaxies in this work have photometry from GALEX. To maintain
consistency we did not use GALEX data to derive host properties.

%Excluding the incomplete GALEX data makes sense. But the argument that UV photometry is overly influenced by star-formation history isn't entirely convincing. One could argue that it's better to have [N,F]UV to understand the AGB star contribution to the redder filters. This turns out to be surprisingly less useful than one would think, but I think the sentence as written in the paper isn't particularly convincing.

\subsection{Separation Between SNe~Ia and Hosts}
We calculate the projected separation between SNe~Ia and their hosts. First, we
compute angular separations between SNe~Ia and the center of their
respective hosts. Then we calculate the distance to each SN~Ia using redshift
and an assumed value of the Hubble-Lema$\rm \hat{i}$tre constant, $\rm H_0 =70 \  km\ sec^{-1}\ Mpc^{-1}$.
With these, we calculate a projected distance in kiloparsecs. 
%Finally, we
%also calculate sizes of galaxies in kiloparsecs from the $r$-band half-light 
%radius,
%and normalize the projected distance by galaxy size which we callNormalized
%Projected Distance ($NPD$).

\section{SN~Ia - Host Correlations}\label{sec:corr}

\subsection{$\Delta m_{15} (B)$ and $s_{BV}$}\label{sec:mor}

\begin{figure*}[htbp]
\begin{center}
\begin{tabular}{cc}
\includegraphics[width=\columnwidth]{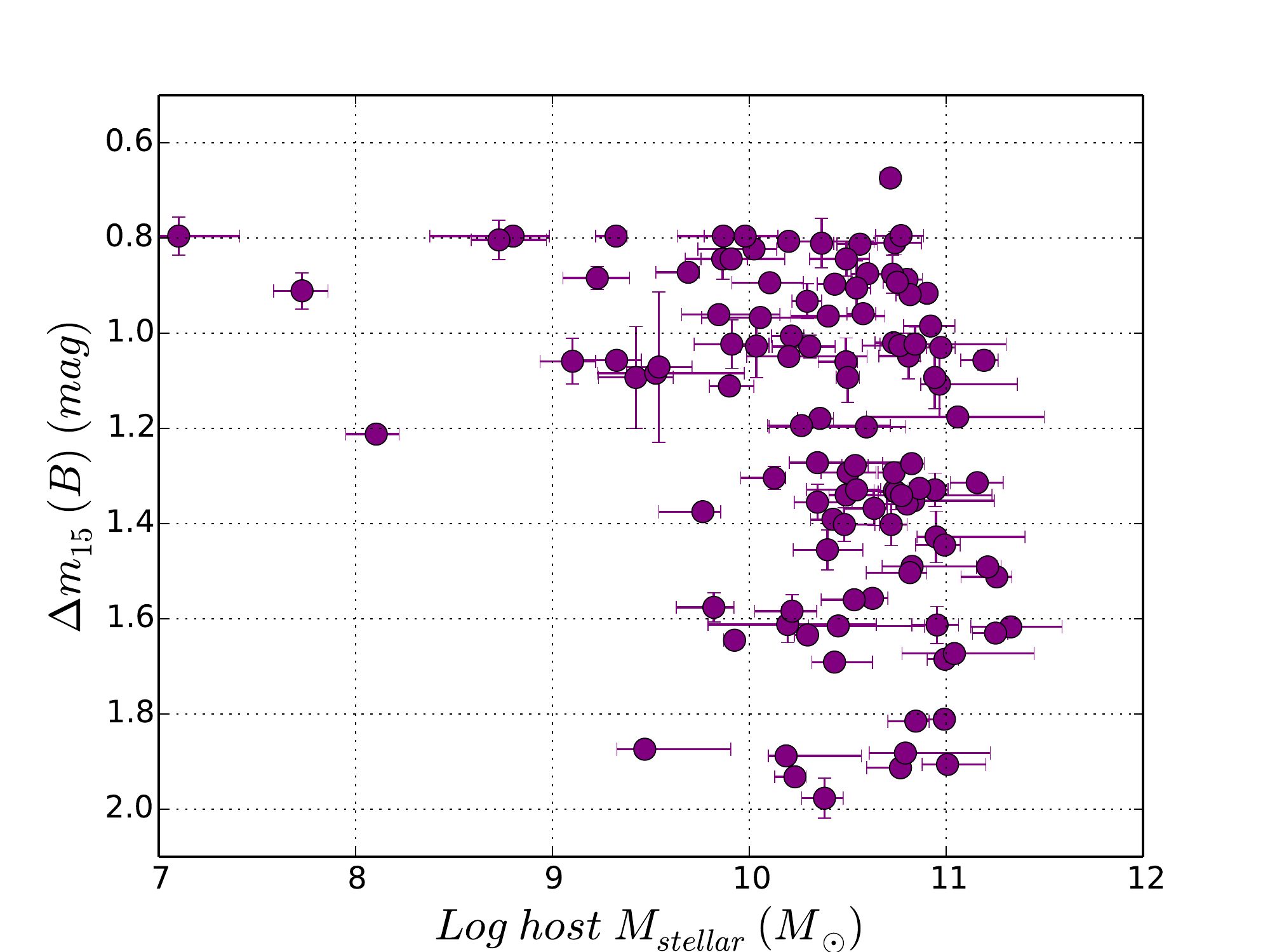} &
\includegraphics[width=\columnwidth]{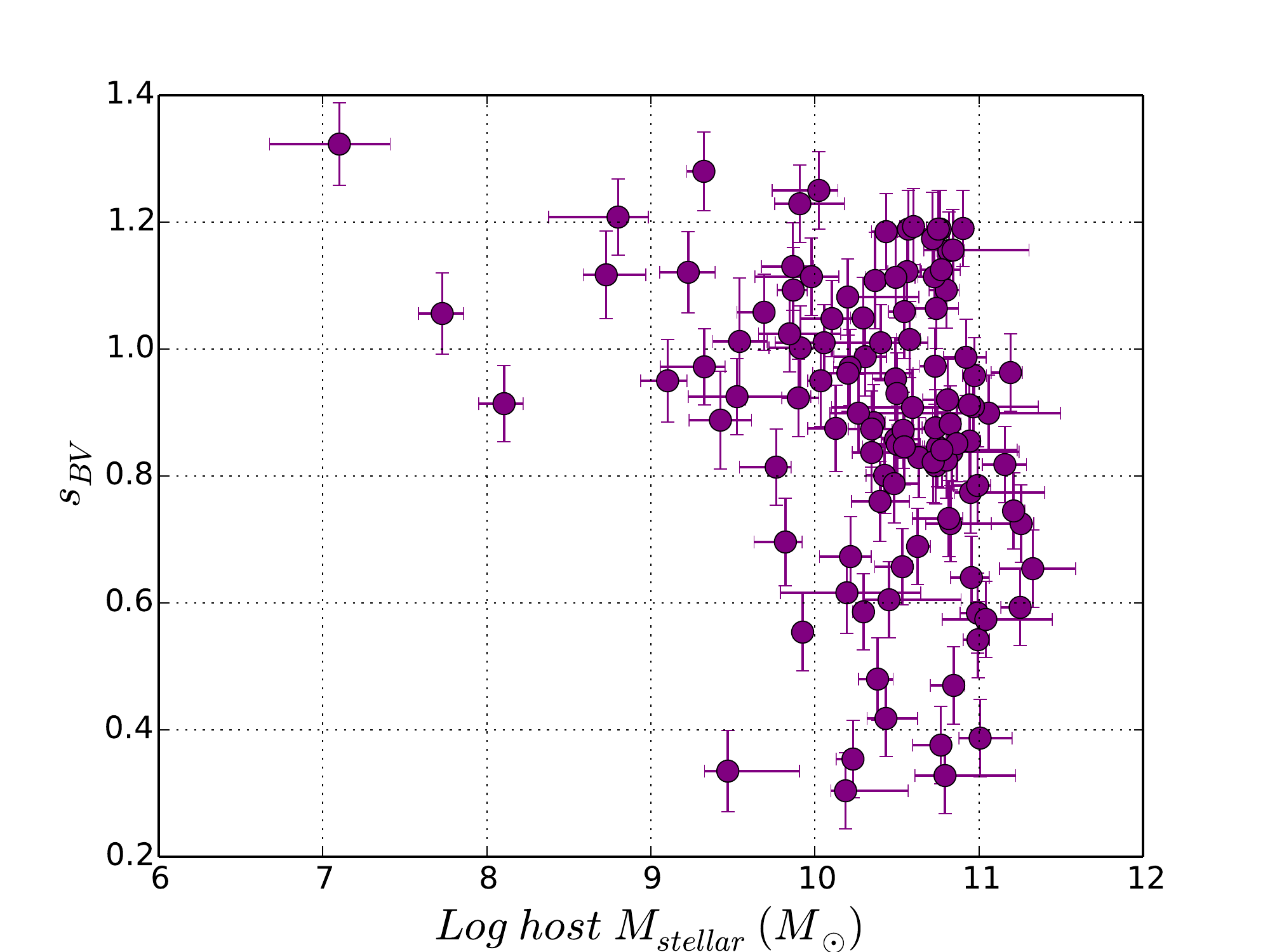} 
\end{tabular}
\caption{Distribution of $\Delta m_{15} (B)$ and $s_{BV}$ with respect to their host stellar mass. See text for discussion. }
\label{corr}
\end{center}
\end{figure*}

\begin{figure*}[ht!]
\begin{center}
\begin{tabular}{cc}
\includegraphics[width=\columnwidth]{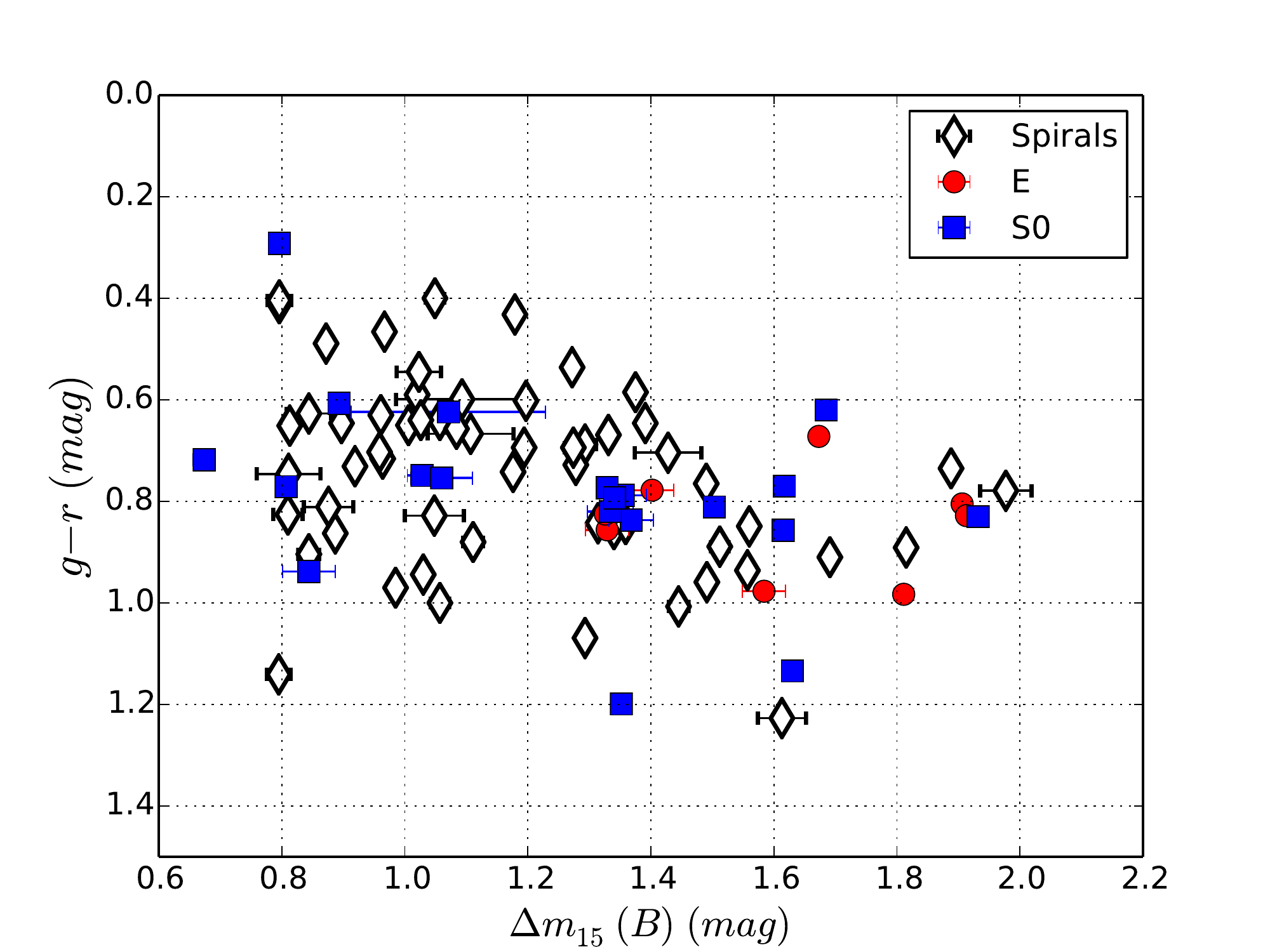} &
\includegraphics[width=\columnwidth]{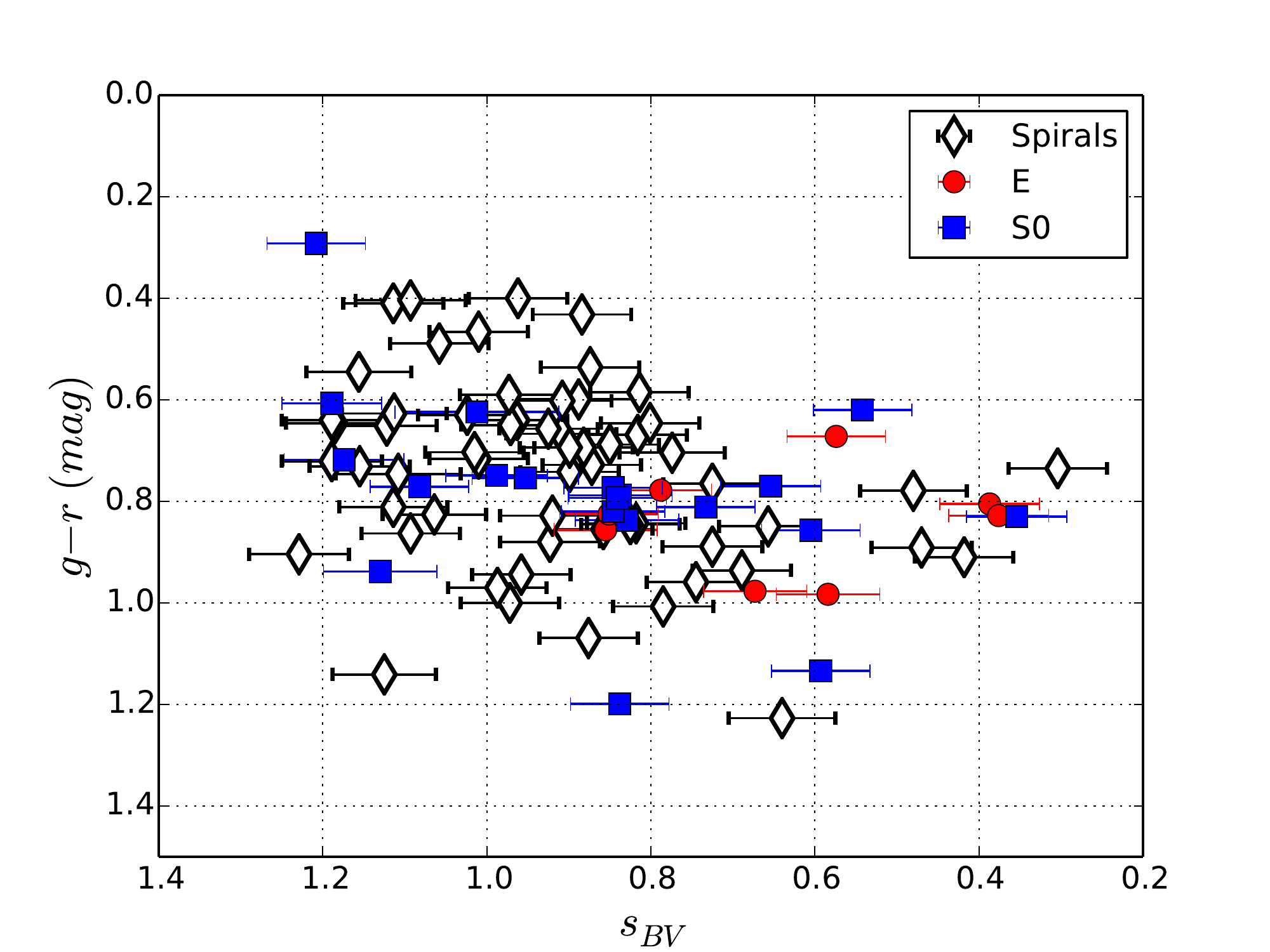} 
\end{tabular}
\caption{Distribution of $\Delta m_{15} (B)$ and $s_{BV}$ with respect to their host $(g-r)$ color. See text for discussion.  }
\label{gr}
\end{center}
\end{figure*}

In this section, we present correlations between SNe~Ia properties and their host galaxies. %We take SN~Ia properties such as $B$-band decline rate, 
%$\Delta m_{15} (B)$\footnote{$\Delta m_{15} (B)$ is defined as the difference in magnitude of    the $B$-band light-curve between maximum and day 15 
%in the rest-frame of the SN~Ia \citep{phillips93}.}, color-stretch parameter, $s_{BV}$, and Hubble residual, $\Delta \mu$, from \citet{burns18}. 
In Fig.~\ref{corr} we show how $B$-band decline rate, $\Delta m_{15} (B)$ and $s_{BV}$ correlate with host stellar mass. $\Delta m_{15} (B)$ is defined as the difference in magnitude of the $B$-band light-curve between maximum and day 15 in the rest-frame of the SN~Ia \citep{phillips93}.
We do not see strong correlations, but as first noted by \citet{sullivan10}, massive hosts are observed to produce SNe~Ia over the full range of decline rates, whereas less massive hosts preferentially give rise to slower declining events. We compare SN~Ia light-curve properties with host galaxy $(g-r)$ color in Fig.~\ref{gr}, where SN~Ia properties are now on the horizontal axis. We reproduce the results in \citet{hamuy00} who found that fast declining SNe~Ia are missing from blue galaxies with $(B-V)\le0.7$ mag, whereas galaxies with $(B-V)\ge 0.7$ mag produce SNe~Ia with a wider range of decline rates. We see a similar pattern in our work, with the division occurring at $(g-r) \sim 0.6$ mag. We also find an excess of SNe~Ia with higher values of $s_{BV}$ (lower values of $\Delta m_{15} (B)$) occurring in bluer galaxies. Finally, we label galaxies with their morphological information obtained from NASA Extragalactic Database (NED)\footnote{http://ned.ipac.caltech.edu}. These classifications are available for $75\%$ of CSP-I SN~Ia  host galaxies.  \citet{hamuy95, hamuy96, hamuy00} found that spiral galaxies preferentially produce slower declining SNe~Ia, whereas elliptical 
galaxies mostly host faster declining events.  In order to test this, we perform a Kolmogorov-Smirnov (K-S) test to study possible intrinsic differences between the two populations of CSP-I SNe~Ia: those exploding in spiral galaxies and those in elliptical or S0 galaxies. In terms of $\Delta m_{15} (B)$, the K-S test gives a value of 0.13 for {\it D-statistics}\footnote{{\it D-statistics} is the maximum difference between two cumulative distributions.} and a {\it p-value} of 0.47.  For this value of {\it D-statistics}, the confidence level, $c(\alpha)$, is smaller than 0.005. Since {\it p-value} $\gg c(\alpha)$, we conclude that the two samples are likely drawn from the same population.

 %\cite{hamuy00} studied host $(B-V)$ color against SN~Ia $B$-band decline rate, and found a lack of fast-declining SNe~Ia among blue galaxies with $(B-V)\le0.7$ mag. 

 %We find that, although spread over, spiral galaxies preferentially host SNe~Ia that have faster decline rates and larger color-stretch parameters. 

\subsection{Hubble Residuals ($\Delta \mu$)}\label{hr}
\begin{figure*}[htbp]
\begin{center}
\includegraphics[width=\textwidth]{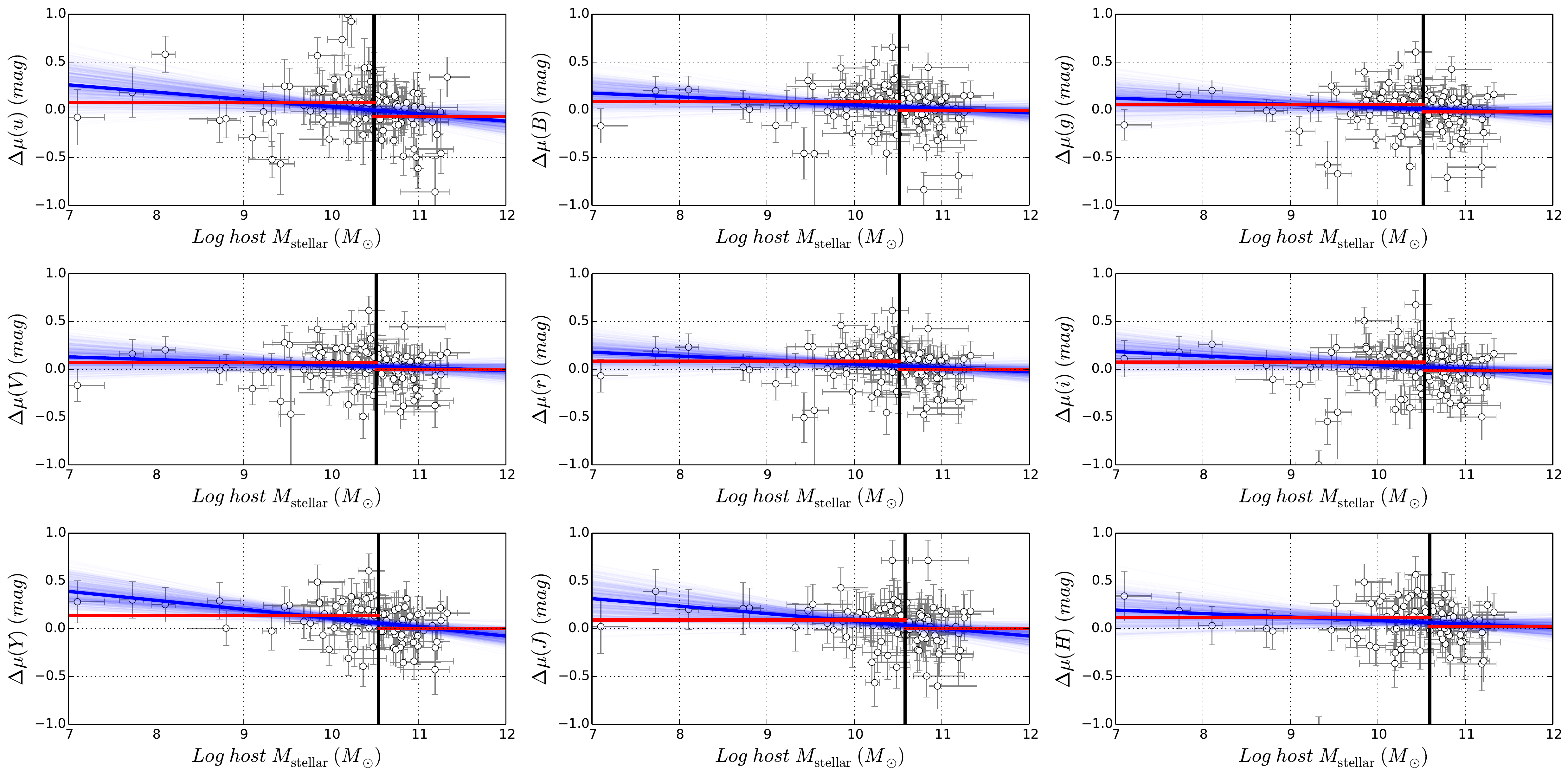} 
\caption{SN~Ia Hubble residual ($\Delta \mu$) vs host stellar mass across $uBgVriYJH$ bands. In each case, the vertical solid black line shows the median split point, the slanted blue thick solid line shows the best-fit linear trend with lighter blue lines contain the 95$\%$ pointwise confidence intervals on the regression line, and red solid lines show the weighted mean of Hubble residuals at both sides of the split point. Slopes of the best-fit lines and Hubble residual offsets are presented in Table~\ref{stat}.}
\label{allres}
\end{center}
\end{figure*}

While the relations between $\Delta m_{15} (B)$ and $s_{BV}$ with host properties are important from an astrophysical point of view (e.g., the explosion mechanism), SN~Ia standardized luminosity variation with host stellar mass is important in deriving cosmological parameters (see \citealt{uddin17b}). In this section, we explore correlations between host stellar mass and SN~Ia Hubble residuals for each of the $uBgVriYJH$ filters employed by the CSP-I, as shown in Fig.~\ref{allres}.
%To state the general conclusion, Fig.~\ref{allres} shows how Hubble residuals of SNe~Ia in $uBgVriYJH$ bands vary with host stellar mass. 

To quantify these correlations, we first calculate slopes of linear fits using a Bayesian method LINMIX \citep{linmix}, which takes into account the possibility of intrinsic scatter in the regression relationship along with possible correlated errors in both dependent and independent variables. Slopes are found to be negative across all filters, implying that SNe~Ia in more massive galaxies are brighter than expected after standardizing the peak magnitudes using the luminosity--decline rate relation. The slope is least negative in $V$-band with a value of $-0.029\pm 0.029$ mag/dex and most negative in $Y$-band  with a value of $-0.093\pm 0.031$ mag/dex. Table~\ref{stat} summarizes the values of the slopes. While the slopes vary from filter to filter, these variations do not deviate significantly from a constant value. %This result indicates that dust is not likely to play a major role in driving SN~Ia Hubble residual- host mass correlations, contrary to what described in \citet{bs20}.

Next, we split our sample into two groups at the median value of host stellar mass, $log \ M_{\rm stellar}(M_{\odot})=10.48$, and compute the weighted mean of Hubble residuals in each group. We find, on average, that SNe~Ia that explode in a host at a higher mass than the median are over-luminous, after light-curve standardization. We call this the Hubble residual offset, $\Delta_{\rm HR}$. The offset varies across the filters, ranging  from $-0.074\pm 0.030$ mag to $-0.147\pm 0.040$ mag. Table~\ref{stat} summarizes the values of $\Delta_{\rm HR}$. Negative values of $\Delta_{\rm HR}$ indicate that SNe~Ia are more luminous in massive hosts. Note that the values of the slopes that we calculate are consistent within the uncertainties with those obtained by \cite{burns18}.

\begin{figure*}[htbp]
\begin{center}
\includegraphics[width=\textwidth]{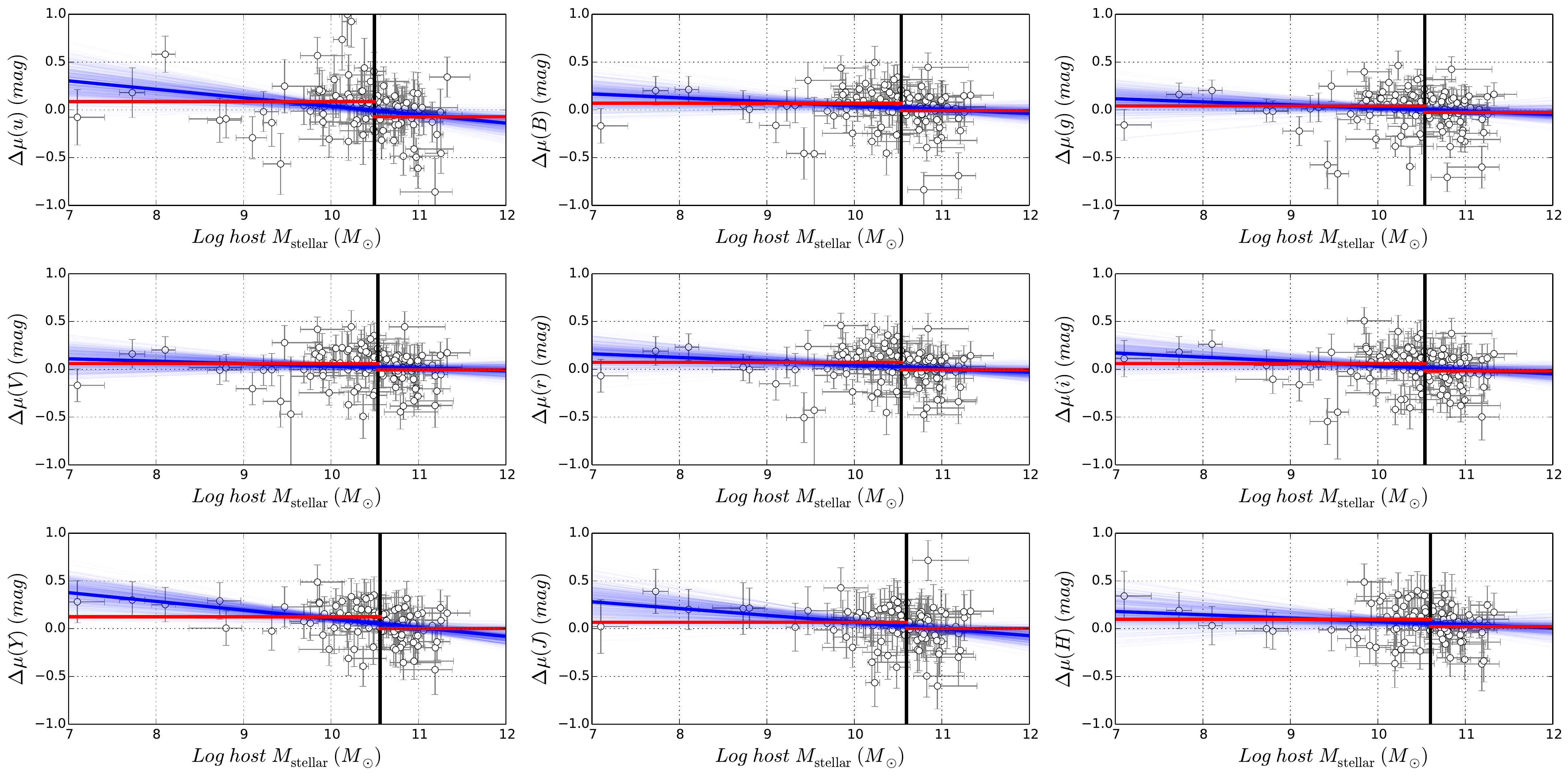} 
\caption{Same as Fig.~\ref{allres} but including only SNe~Ia with $z>0.005$.}
\label{allresz1}
\end{center}
\end{figure*}

\begin{figure*}[htbp]
\begin{center}
\includegraphics[width=\textwidth]{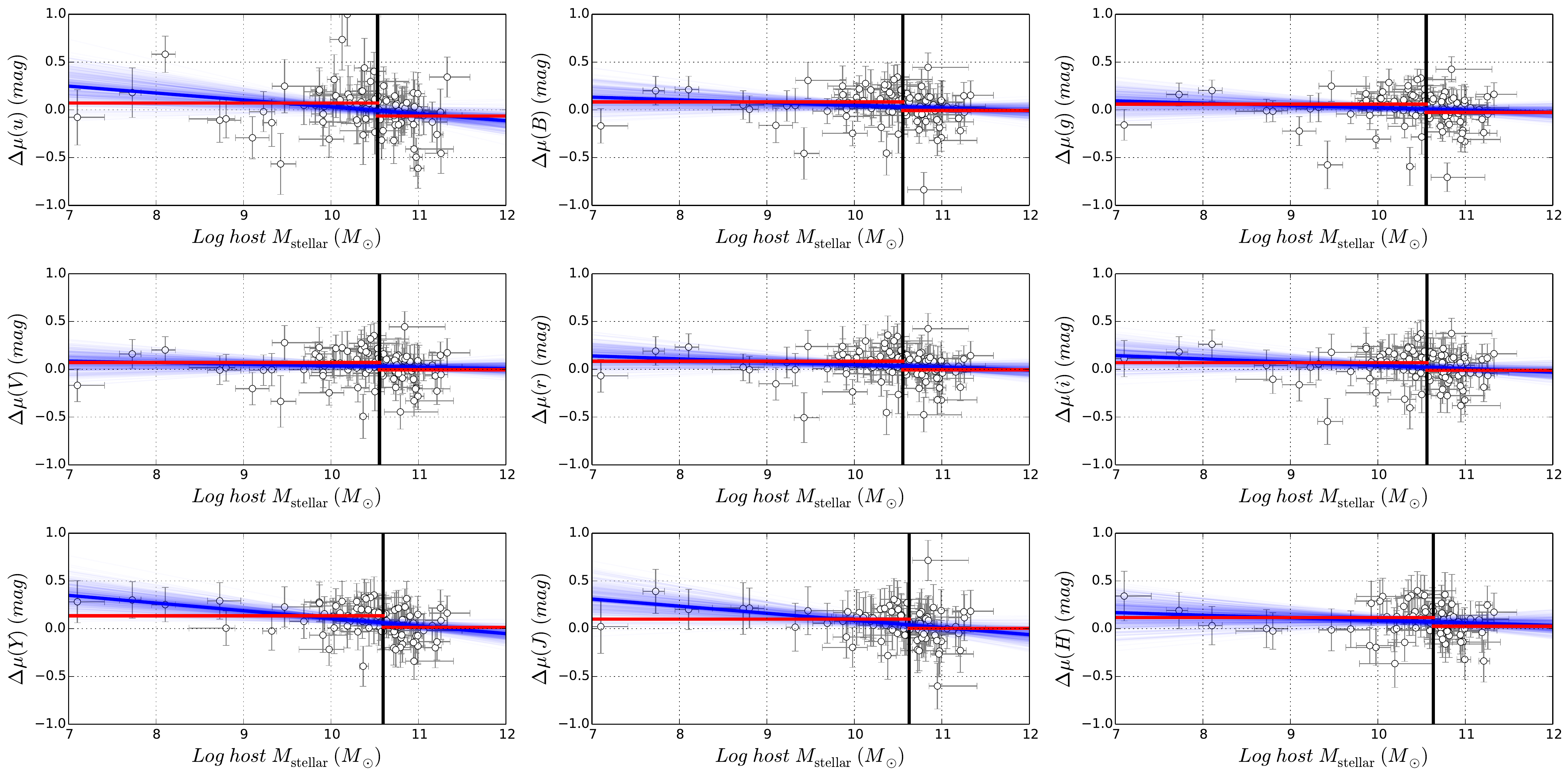} 
\caption{Same as Fig.~\ref{allres} but including only SNe~Ia with $z>0.01$.}
\label{allresz2}
\end{center}
\end{figure*}

\begin{figure*}[htbp]
\begin{center}
\begin{tabular}{cc}
\includegraphics[width=\columnwidth]{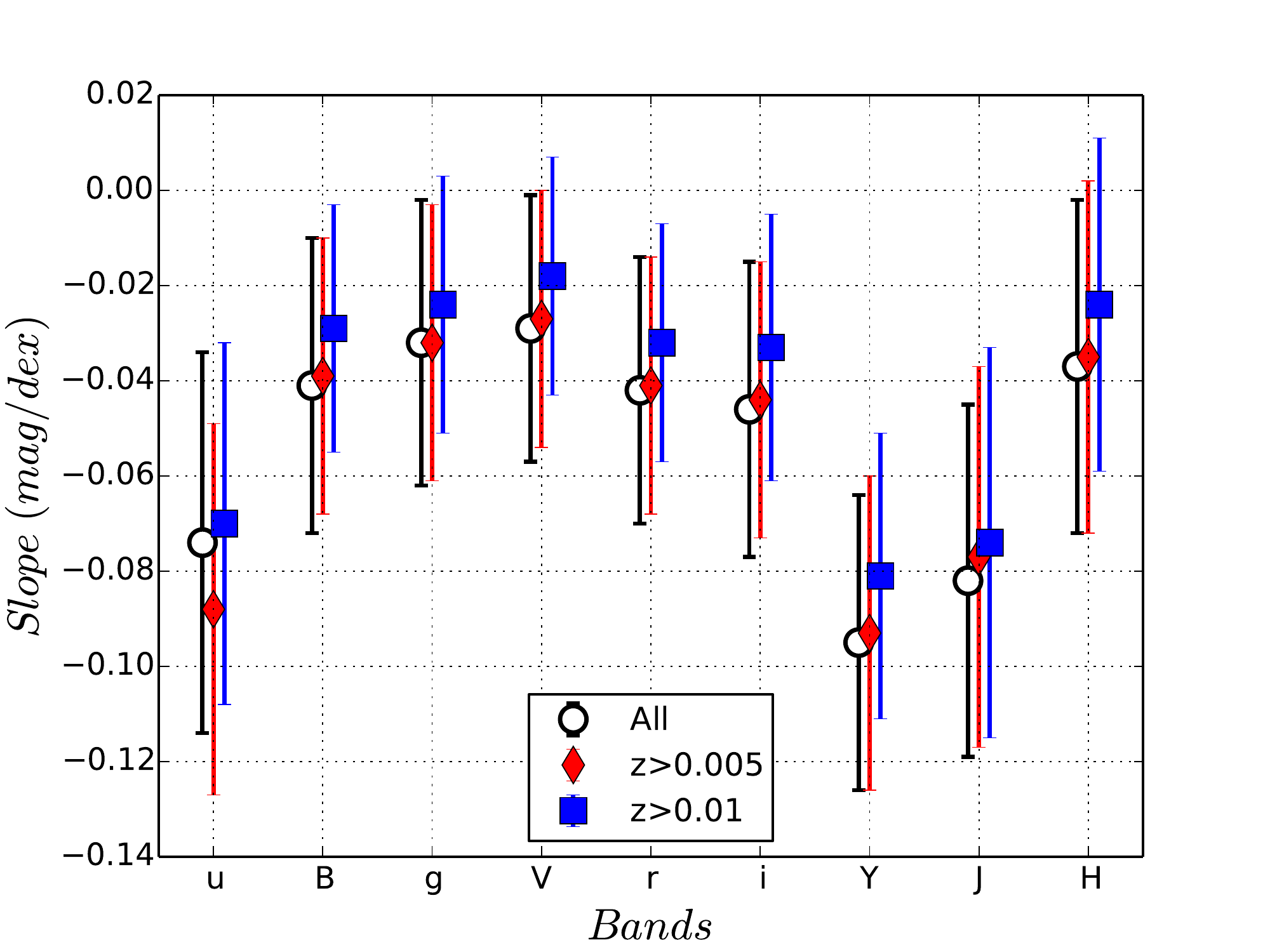} &
\includegraphics[width=\columnwidth]{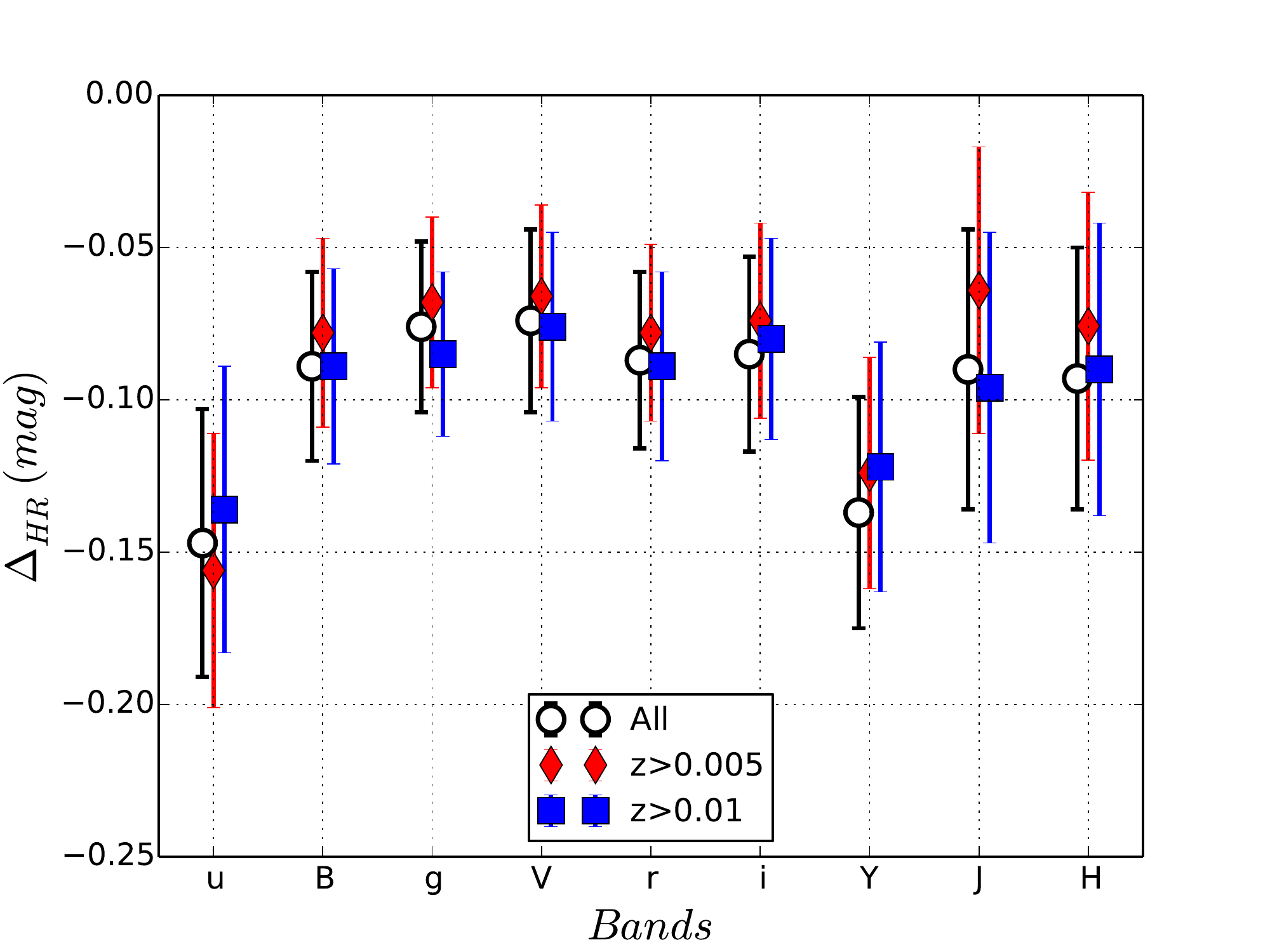} 
\end{tabular}
\caption{Variation of slope (left) and $\Delta_{\rm HR}$ (right) in different bands. They are shown for the parent sample, $z>0.005$ sample, and $z>0.01$ sample. See Table~\ref{stat} for the values of slopes and $\Delta_{\rm HR}$.}
\label{corrz}
\end{center}
\end{figure*}

SNe~Ia in the nearby universe are affected by peculiar velocities, which in turn affects both SN~Ia luminosities and host galaxy stellar mass in a correlated manner. We might, therefore, expect these correlations to change when nearby SNe~Ia are excluded.  Figures~\ref{allresz1} and~\ref{allresz2} display the correlations after excluding SNe~Ia with redshifts $z<0.005 $ and $z<0.01$, respectively. Table~\ref{stat} summarizes the slopes and $\Delta_{\rm HR}$ values obtained in different filters after applying these redshift cuts, with the results displayed visually in Fig.~\ref{corrz}. We find that the slopes shift towards positive values, but these shifts are not significant. The $\Delta_{\rm HR}$ values change even less after excluding nearby SNe~Ia. 

\begin{figure*}[htbp]
\begin{center}
\includegraphics[width=\textwidth]{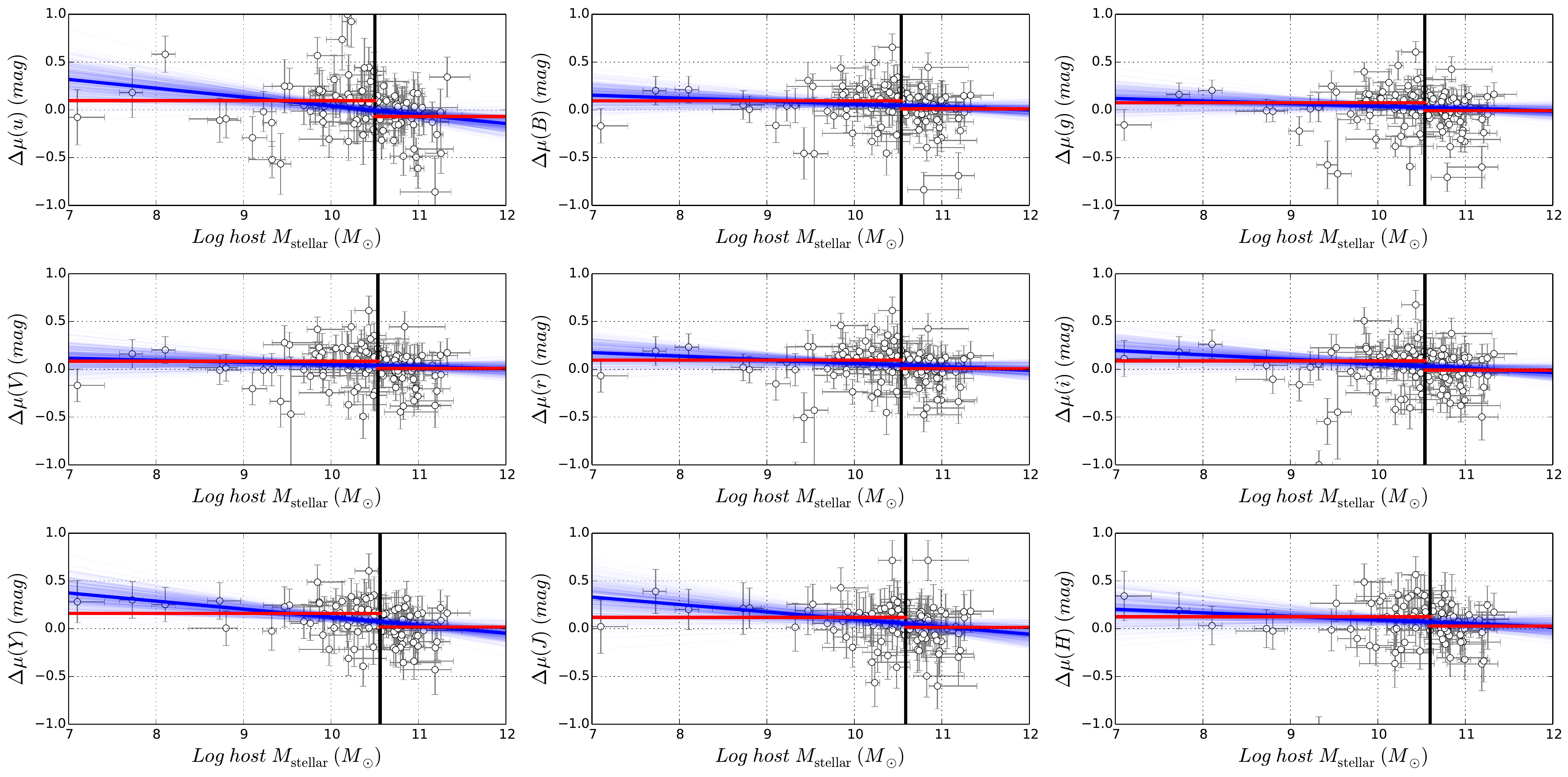} 
\caption{Same as Fig.~\ref{allres} but including only SNe~Ia with $s_{BV}>0.5$ and $E(B-V)<0.5$.}
\label{allress}
\end{center}
\end{figure*}

Finally, we investigate two special cases. Typically, fast decliners and heavily reddened SNe~Ia are not included in most cosmological analyses. To test how the SN~Ia Hubble residual-host mass correlations might change by excluding these objects, we eliminate all SNe in the CSP-I sample with $s_{BV}>0.5 $ and $E(B-V)<0.5$~mag. The resulting sample is plotted in Fig.~\ref{allress} and slopes and $\Delta_{\rm HR}$ are included in Table~\ref{stat}. We find small shifts in the values of the slopes and $\Delta_{\rm HR}$, but these are consistent within the errors with the values obtained for the full sample. 

\begin{figure*}[htbp]
\begin{center}
\includegraphics[width=\textwidth]{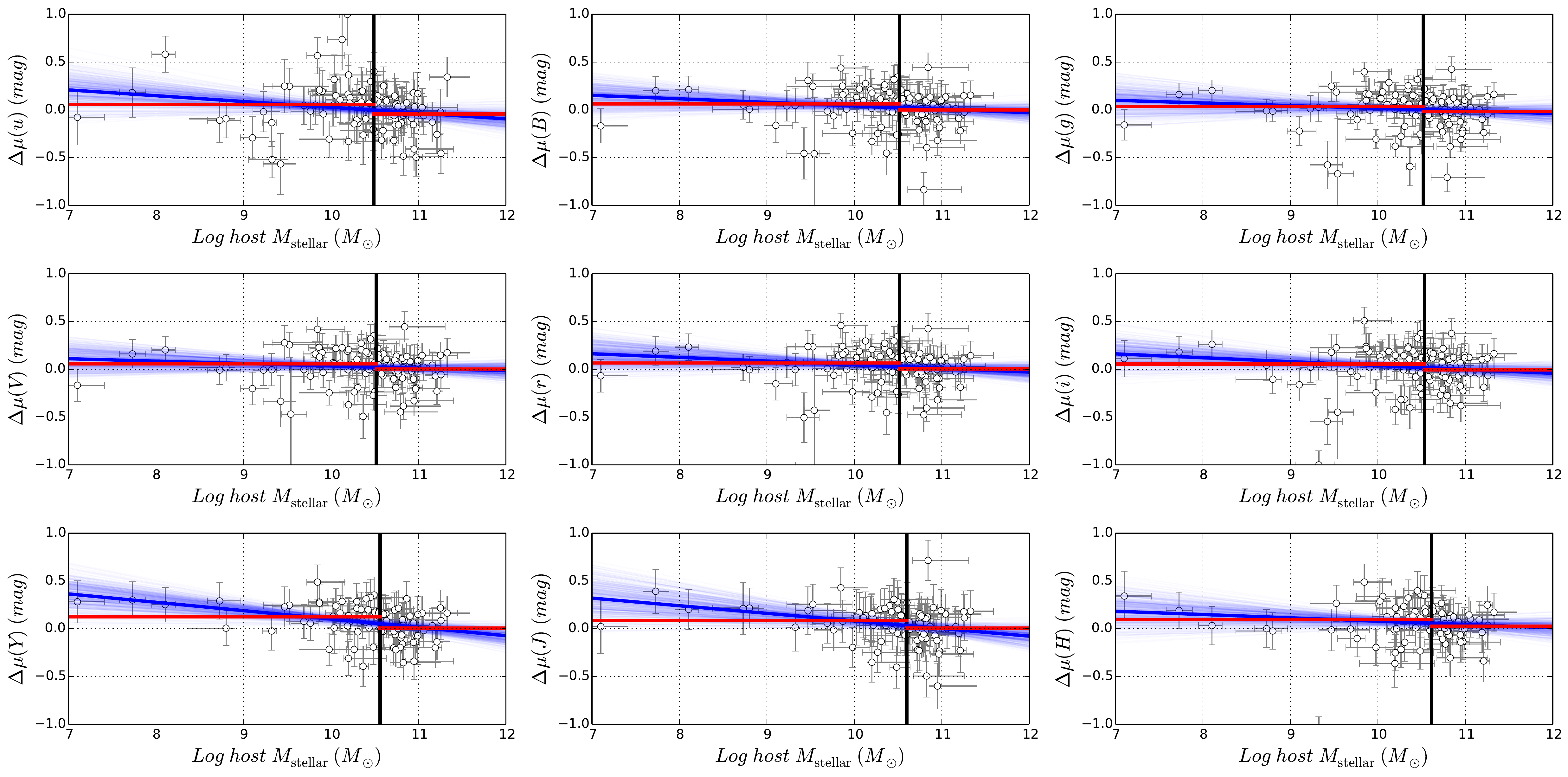} 
\caption{Same as Fig.~\ref{allres} but excluding 91T- and 91bg-like SNe~Ia.}
\label{allres91}
\end{center}
\end{figure*}

The CSP-I sample includes several SNe identified as 91T- and 91bg-like events. These classification are from \citet{folatelli13}, and are based on spectral comparison with SNID templates \citep{blondin07}. The Hubble residual-host mass correlations without these objects are plotted in Fig.~\ref{allres91}, and the resulting slopes and $\Delta_{\rm HR}$ values are included in Table~\ref{stat}.  Again, we find only small shifts in these values. 

It is important to note that the correlation between SN~Ia Hubble residual and host galaxy stellar mass exists across all filters, \textit{from optical to near-infrared wavelengths}. This implies that differences
in the properties of the dust in low- and high-mass host galaxies is not likely the driving factor in these correlations, contrary to what is postulated by \citet{bs20}.

\begin{figure}[htbp]
\begin{center}
\includegraphics[width=\columnwidth]{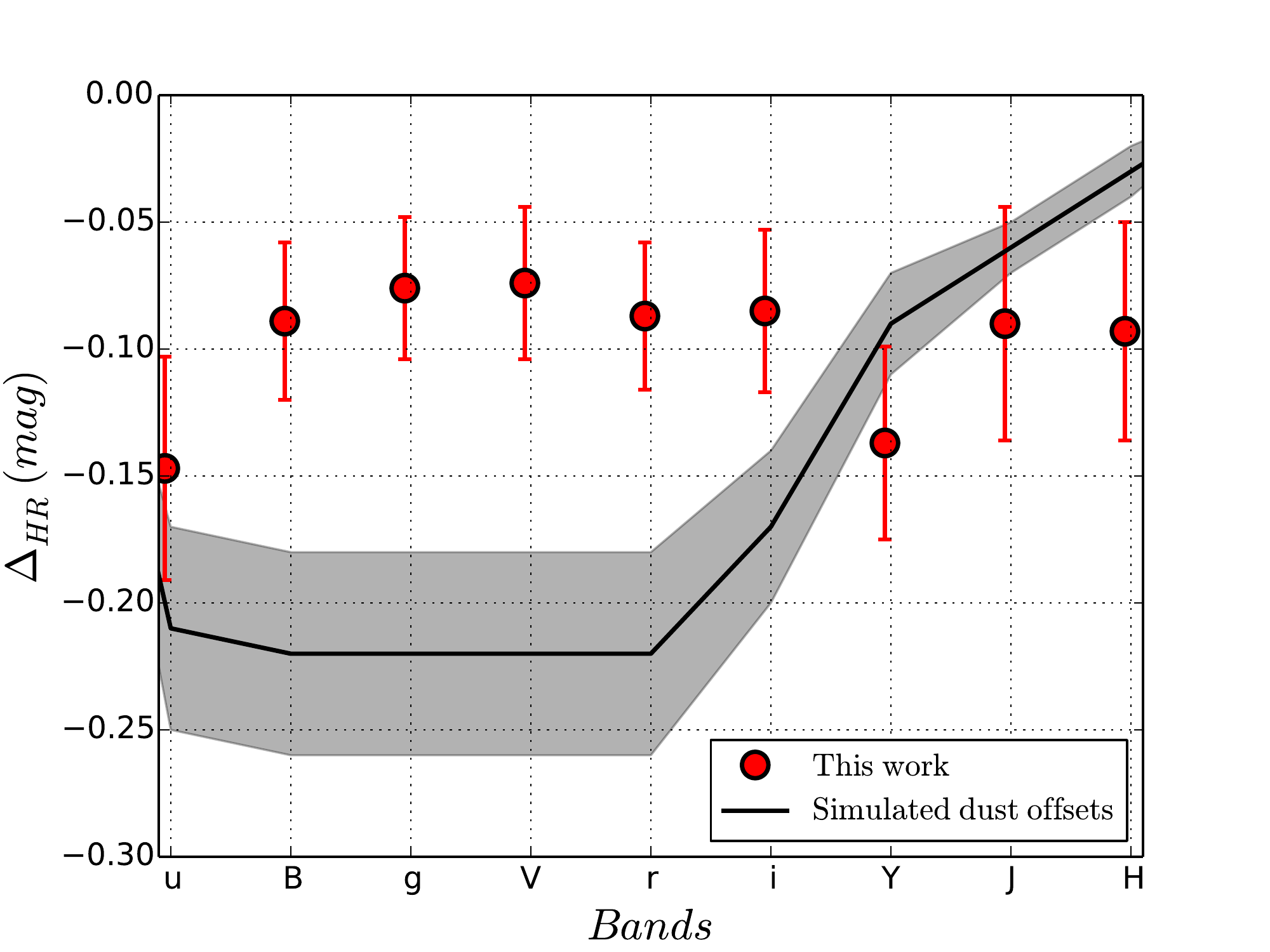} 
\caption{Similar to Fig.~\ref{corrz} (right) but adding simulated dust offsets according to \citet{bs20}. See \autoref{hr} for discussion.}
\label{offsim}
\end{center}
\end{figure}

We further explore this in Fig.~\ref{offsim} where we plot the wavelength-dependence of Hubble residuals one would expect from the \citet{bs20} model. These were estimated by randomly drawing a set of 100 color excesses, $E(B-V)_{\rm host}$, and values of the ratio of total to selective absorption, $R_V^{\rm host}$ , for a low-mass and high-mass sample consistent with the findings of \citet{bs20}. For $E(B-V)_{\rm host}$ we drew from an exponential distribution with scale $\tau = 0.18$ mag for the high-mass sample and $\tau = 0.16$ mag for the low-mass sample. For $R_V^{\rm host}$ , we drew from a normal distribution with a mean $R_V^{\rm host} = 1.5$ and standard deviation $\sigma_{R_V^{\rm host}} = 0.8$ for the high-mass sample, and a mean $R_V^{\rm host}= 2.5$  and $\sigma_{R_V^{\rm host}} = 2.2$ for the low-mass sample. From these we computed, for each filter, the mean offset in luminosity that would result from assuming (incorrectly) there was a single average $R_V^{\rm host}$  for both samples.  As is seen in Fig.~\ref{offsim}, the resulting prediction for the behavior of the Hubble residual as a function of wavelength is not a good fit to the CSP-I observations.

%The slope of the SN~Ia Hubble residual-host mass correlation seems to change, but this change is not significant. The change in $\Delta_{\rm HR}$ is also insignificant. 

%table
\begin{table}[ht]
\footnotesize
\caption{Statistics on the SN~Ia Hubble residual-host galaxy stellar mass correlation. Negative values in slopes indicate anti-correlation, i.e., SNe~Ia are more luminous as host mass increases (see Fig.~\ref{allres}, \ref{allresz1}, and \ref{allresz2}). Errors in slopes and $\Delta_{\rm HR}$ are in parentheses.}
%\begin{center}
\begin{tabular}{cccc}
\hline
\hline
Sample & Filter & Slope & $\Delta_{\rm HR}$ \\
& &  (mag~dex$^{-1}$) & (mag)\\
\hline
All  &$u$ &-0.076 (0.041)& -0.147 (0.044)\\
& $B$ &-0.041 (0.031)& -0.089 (0.031)\\
& $g$ &-0.032 (0.031)& -0.076 (0.028)\\
& $V$ &-0.029 (0.029)& -0.074 (0.030)\\
& $r$ &-0.041 (0.028)& -0.087 (0.029)\\
& $i$ &-0.046 (0.030)& -0.085 (0.032)\\
& $Y$ &-0.093 (0.031)& -0.137 (0.038)\\
& $J$ &-0.078 (0.040)& -0.090 (0.046)\\
& $H$ &-0.036 (0.036)& -0.093 (0.043)\\

\hline
$z>0.005$ & $u$ &-0.088 (0.040)& -0.156 (0.045)\\
& $B$ &-0.039 (0.030)& -0.078 (0.031)\\
& $g$ &-0.031 (0.029)& -0.068 (0.028)\\
& $V$ &-0.026 (0.027)& -0.066 (0.030)\\
& $r$ &-0.039 (0.027)& -0.078 (0.029)\\
& $i$ &-0.045 (0.029)& -0.074 (0.032)\\
& $Y$ &-0.090 (0.032)& -0.124 (0.038)\\
& $J$ &-0.076 (0.042)& -0.064 (0.047)\\
& $H$ &-0.039 (0.033)& -0.075 (0.044)\\
\hline
$z>0.01$ & $u$ &-0.072 (0.038)& -0.136 (0.047)\\
& $B$ &-0.029 (0.027)& -0.089 (0.032)\\
& $g$ &-0.024 (0.027)& -0.085 (0.027)\\
& $V$ &-0.016 (0.024)& -0.076 (0.031)\\
& $r$ &-0.030 (0.025)& -0.089 (0.031)\\
& $i$ &-0.034 (0.026)& -0.080 (0.033)\\
& $Y$ &-0.080 (0.031)& -0.122 (0.041)\\
& $J$ &-0.074 (0.037)& -0.096 (0.051)\\
& $H$ &-0.027 (0.035)& -0.090 (0.048)\\

\hline
$s_{BV}>0.5$ & $u$ &-0.092 (0.038)& -0.166 (0.046)\\
$\&$ & $B$ &-0.031 (0.026)& -0.086 (0.031)\\
$E(B$-$V)_{\rm host}<0.5$& $g$ &-0.027 (0.025)& -0.083 (0.025)\\
& $V$ &-0.022 (0.026)& -0.077 (0.031)\\
& $r$ &-0.037 (0.027)& -0.090 (0.031)\\
& $i$ &-0.046 (0.027)& -0.096 (0.033)\\
& $Y$ &-0.084 (0.032)& -0.142 (0.042)\\
& $J$ &-0.078 (0.039)& -0.103 (0.050)\\
& $H$ &-0.037 (0.036)& -0.097 (0.047)\\

\hline
Exclude  91T-  & $u$ &-0.061 (0.038)& -0.099 (0.045)\\
$\&$& $B$ &-0.036 (0.030)& -0.063 (0.031)\\
91bg-like SNe Ia & $g$ &-0.028 (0.029)& -0.051 (0.028)\\
& $V$ &-0.025 (0.028)& -0.052 (0.031)\\
& $r$ &-0.038 (0.028)& -0.061 (0.030)\\
& $i$ &-0.041 (0.029)& -0.060 (0.033)\\
& $Y$ &-0.087 (0.033)& -0.118 (0.040)\\
& $J$ &-0.080 (0.039)& -0.080 (0.049)\\
& $H$ &-0.035 (0.036)& -0.067 (0.046)\\
\hline
\end{tabular}
%\end{center}
\label{stat}
\end{table}%

\normalsize

\subsection{Effect of SN~Ia-Host Separation}

\begin{figure*}[htbp]
\begin{center}
\includegraphics[width=\textwidth]{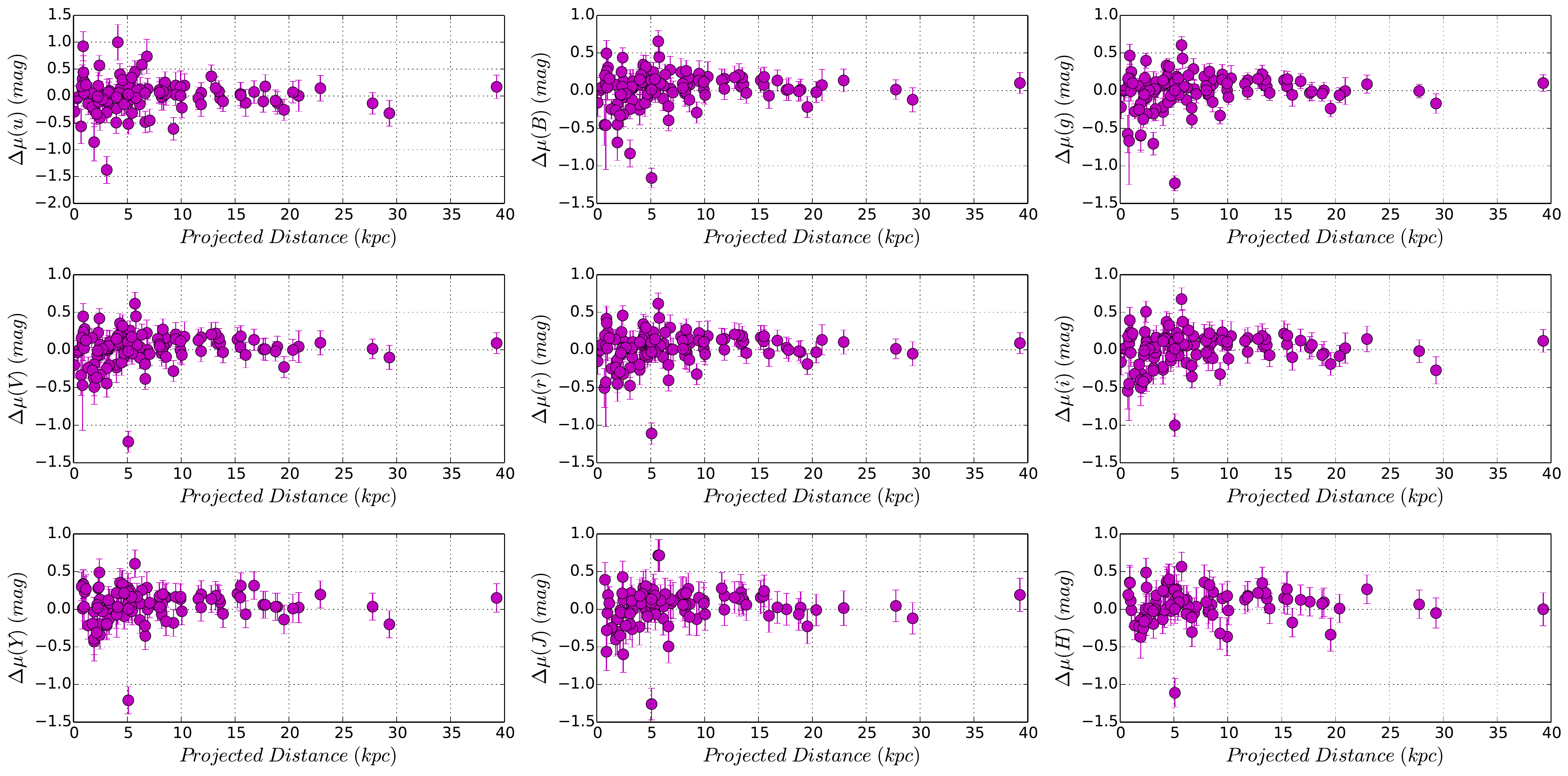} 
\caption{ Variation of $\Delta \mu$ with projected distance from host centers in $uBgVriYJH$ bands. In all cases $\Delta \mu$ have smaller dispersion when they explode farther from their host centers.}
\label{sepall}
\end{center}
\end{figure*}

We also investigate how Hubble residual varies as a function of the positions of the SNe in their host galaxies. This is displayed in Fig.~\ref{sepall} across the $uBgVriYJH$ filters.   In Table~\ref{scd}, values of the dispersion in Hubble residuals for SNe~Ia within 10 kpc of their host centers and for those beyond 10 kpc are tabulated for each filter. We find that SNe~Ia exploding farther than 10 kpc from their host centers have a dispersion in Hubble residuals that is smaller than that for the SNe that explode within 10 kpc of their host centers. The reduction in the dispersion for the full sample varies from $\rm 50\% \ to \ 60\%$ across the filters.  For the $B$ band, which is representative of all the filters, the dispersion in Hubble residual for the SNe~Ia beyond 10 kpc projected distance is $\sim 0.17$~mag smaller than the dispersion for
the SNe~Ia within 10 kpc.

\begin{table}[htp]
\caption{Dispersion in Hubble residuals of SNe Ia that are within 10 kpc from their host centers and those that are beyond 10 kpc in various bands.}
\begin{center}
\begin{tabular}{cccc}
\hline
\hline
Filter & Dispersion in $\Delta \mu \ (mag)$ &\\
\hline
 & $Distance <10 \ kpc$  & $Distance >10 \ kpc$  \\
\hline
All & &\\
\hline
$ u $ & 0.34 & 0.16\\
$ B $ & 0.27 & 0.10\\
$ g $ & 0.27 & 0.11\\
$ V $ & 0.25 & 0.10\\
$ r $ & 0.25 & 0.10\\
$ i $ & 0.25 & 0.12\\
$ Y $ & 0.25 & 0.12\\
$ J $ & 0.28 & 0.12\\
$ H $ & 0.25 & 0.15\\
\hline
$z>0.005$ &&\\
\hline
$ u $ & 0.34 & 0.16\\
$ B $ & 0.26 & 0.10\\
$ g $ & 0.27 & 0.11\\
$ V $ & 0.24 & 0.10\\
$ r $ & 0.24 & 0.10\\
$ i $ & 0.24 & 0.12\\
$ Y $ & 0.25 & 0.12\\
$ J $ & 0.27 & 0.12\\
$ H $ & 0.24 & 0.15\\
\hline
$z>0.01$ &&\\
\hline
$ u $ & 0.31 & 0.14\\
$ B $ & 0.20 & 0.10\\
$ g $ & 0.20 & 0.11\\
$ V $ & 0.18 & 0.10\\
$ r $ & 0.18 & 0.10\\
$ i $ & 0.18 & 0.12\\
$ Y $ & 0.17 & 0.13\\
$ J $ & 0.20 & 0.12\\
$ H $ & 0.18 & 0.15\\
\hline
\end{tabular}
\end{center}
\label{scd}
\end{table}%

%We see that SNe~Ia that explode farther away from host centers are more uniform than those that explode closer to their host centers.
We perform a Kolmogorov-Smirnov (K-S) goodness-to-fit test to study possible intrinsic differences between these two populations, one within 10 kpc projected distance, and the other beyond 10 kpc projected distance. The K-S test gives a value of 0.257 for {\it D-statistics} and a {\it p-value} of 0.218.  For this value of {\it D-statistics}, the confidence level, $c(\alpha)$, is smaller than 0.001. Since {\it p-value} $\gg c(\alpha)$, we conclude that the two samples are likely drawn from the same population.

%accept the null hypothesis of the K-S test, and conclude that these two distributions are drawn from the same parent distribution.

We note that there are 20 SNe~Ia beyond 10 kpc from their host centers, perhaps causing an extra undertainty due to small number statistics. To check this possible uncertainty and verify our finding, we perform a Monte Carlo simulation. The steps are: 1.) we first calculate the standard deviation of Hubble residuals for 20 SNe~Ia that are beyond 10 kpc from their host centers; 2.) we randomly draw 20 SNe~Ia Hubble residuals from SNe~Ia that are within 10 kpc from their host centers and calculate the standard deviation; and 3.) we compare the standard deviations of the Hubble residuals found in steps 1 and 2.  After running a million Monte Carlo realizations in this manner, the mean of the differences in the standard deviations is found to be $0.12\pm0.03$ mag. This implies that the result we have found did not occur by chance. 

%To check if the peculiar velocities of nearby SNe~Ia have an effect, we recalculate the dispersions in Hubble residual for SNe~Ia at $z>0.005$ and at $z>0.01$ (see Table~\ref{scd}). For the bluer filters, we still find a significant difference in the dispersion between SNe~Ia within 10 kpc versus those beyond 10 kpc from the centers of their host, but with the dispersions becoming more similar at longer wavelengths. 

%To check if the peculiar velocities of nearby SNe~Ia have an effect, we recalculate the dispersions in Hubble residual for SNe~Ia at $z>0.005$ and at $z>0.01$ (see Table~\ref{scd}). For SNe~Ia that explode beyond $z>0.005$, there is no appreciable change in their dispersions across the wavelengths. But beyond $z>0.01$, we notice $\sim0.05$ mag change in the dispersion for SNe~Ia within 10 kpc from the centers of their host. Finally, beyond $z>0.01$, and for the bluer filters, we still find a significant difference in the dispersion between SNe~Ia within 10 kpc versus those beyond 10 kpc from the centers of their host, but the dispersions become more similar at longer wavelengths. This implies that individually or together dust and peculiar velocity are driving the observed difference in dispersion for SNe~Ia within 10 kpc versus those beyond 10 kpc from the centers of their host.

To check if the peculiar velocities of nearby SNe~Ia have an effect, we recalculate the dispersions in Hubble residual for SNe~Ia at $z>0.005$ and at $z>0.01$ (see Table~\ref{scd}). For SNe~Ia at redshifts $z >0.005$ there is no appreciable change in their dispersions across the wavelengths compared to the full sample. But if we limit the sample to those at redshifts $z>0.01$, a $\sim0.07$ mag decrease in the dispersions for SNe~Ia within 10 kpc is observed in all filters except $u$. Note, also, that the differences in the dispersions for the SNe beyond 10 kpc compared to those within 10 kpc grow smaller and tend to decrease with wavelength.  This suggests that a combination of peculiar velocities and small errors in host dust corrections may be responsible for the observed difference in dispersion for SNe~Ia within 10 kpc versus those beyond 10 kpc from the centers of their hosts.

Finally, looking at the host galaxy mass distribution, we find that for the inner sample ($< 10$ kpc), 75$\%$ of the hosts are massive ($log \ M_{\rm stellar}(M_{\odot}) >10$), whereas for the outer sample 
($> 10$ kpc), 88$\%$ of the hosts are massive. Hence, there is a small excess (13$\%$) of massive host galaxies for SNe~Ia that explode beyond 10 kpc from their host centers.

%\subsection{Ejecta Velocities}
%\begin{figure*}[ht]
%\begin{center}
%\begin{tabular}{cc}
%\includegraphics[width=.9\columnwidth]{CSPI_hk_mass.pdf} &
%\includegraphics[width=.9\columnwidth]{CSPI_ir_mass.pdf}\\
%\includegraphics[width=.9\columnwidth]{CSPI_4130_mass.pdf} &
%\includegraphics[width=.9\columnwidth]{CSPI_5449_mass.pdf}\\ 
%\includegraphics[width=.9\columnwidth]{CSPI_5622_mass.pdf} &
%\includegraphics[width=.9\columnwidth]{CSPI_5972_mass.pdf}\\ 
%\includegraphics[width=.9\columnwidth]{CSPI_6355_mass.pdf} &
%\includegraphics[width=.9\columnwidth]{CSPI_dv20_mass.pdf}
%\end{tabular}
%\caption{}
%\label{ejecta}
%\end{center}
%\end{figure*}

\section{Discussion}\label{sec:dis}

\begin{enumerate}[label=(\alph*)]

\item Our results on the correlation between SN~Ia Hubble residual and host stellar mass are consistent with other published studies. When we compare our results with previous studies we find that results vary in magnitude (and significance) due to different sample size and analysis. Except for \citet{burns18}, all previous studies used $B$-band Hubble residuals. \cite{pan14} studied the SN~Ia Hubble residual-host mass correlation with a sample of 50 SNe~Ia from the Palomar Transient Factory. They found a value of $\Delta_{\rm HR}$ of $-0.08  \pm 0.04$ mag, and a slope of $-0.041\pm 0.030$ mag~dex$^{-1}$. \cite{childress13} used 115 SNe~Ia from the Nearby Supernova Factory program and found a $\Delta_{\rm HR}$ of $-0.085\pm 0.028$ mag, and a slope of $-0.043\pm 0.014$ mag~dex$^{-1}$. Similar and consistent results were also presented from the SDSS (\citealt{lampeitl10a}) and SNLS (\citealt{sullivan10}) samples. \citet{uddin17b} combined both spectroscopically confirmed and photometrically classified SNe~Ia from multiple surveys and undertook a consistent analysis. They found a $\Delta_{\rm HR}$ of $-0.05\pm 0.01$ mag from a sample of 1338 SNe~Ia. In the CSP-I sample, we find similar correlations between Hubble residuals and host stellar mass across all wavelengths from optical to near-infrared, in apparent contradiction to the suggestion that dust explains the luminosity correction \citep{bs20}.  

\item \cite{galbany12} found no correlation ($\rm -0.0019\pm 0.0022  \ mag \ kpc^{-1}$) between SN~Ia Hubble residual and projected distance. The study presented in \citet{hicken09} is closer to this work, where they found 0.05 mag less dispersion in SNe~Ia exploding beyond 12 kpc from their host centers. 

Why do SNe~Ia have larger Hubble residuals when they explode closer to their hosts, while the dispersion gets smaller when they explode farther away? We suspect that dust in host galaxies may play a role here, particularly since the difference in dispersion is largest at bluer wavelengths after eliminating the nearest SNe~Ia. We examine this in Fig~\ref{ebv}, where we encode $E(B-V)_{\rm host}$ information in the plot of Hubble residual vs. projected distance. We see that SNe~Ia that explode beyond 10 kpc from host centers have a smaller variation in $E(B-V)_{\rm host}$. On the other hand, SNe~Ia that explode within 10 kpc of host centers have a large variation of $E(B-V)_{\rm host}$. Moreover, it seems that SNe~Ia with higher color excess are also more luminous after light-curve standardization (negative values of Hubble residuals).  

%It will be interesting to see how dust is distributed along the SN~Ia-host separation
In Fig.~\ref{rv}, we plot both $E(B-V)_{\rm host}$ and the extinction law $R_V^{\rm host}$ with respect to the projected distance.
One can see that SNe~Ia that explode within 10 kpc of their host centers have large variations in color excess, whereas those that explode beyond 10 kpc of their host center have smaller variations in color excess. The extinction law seems not to vary according to separation.

\begin{figure}[htbp]
\begin{center}
\includegraphics[width=\columnwidth]{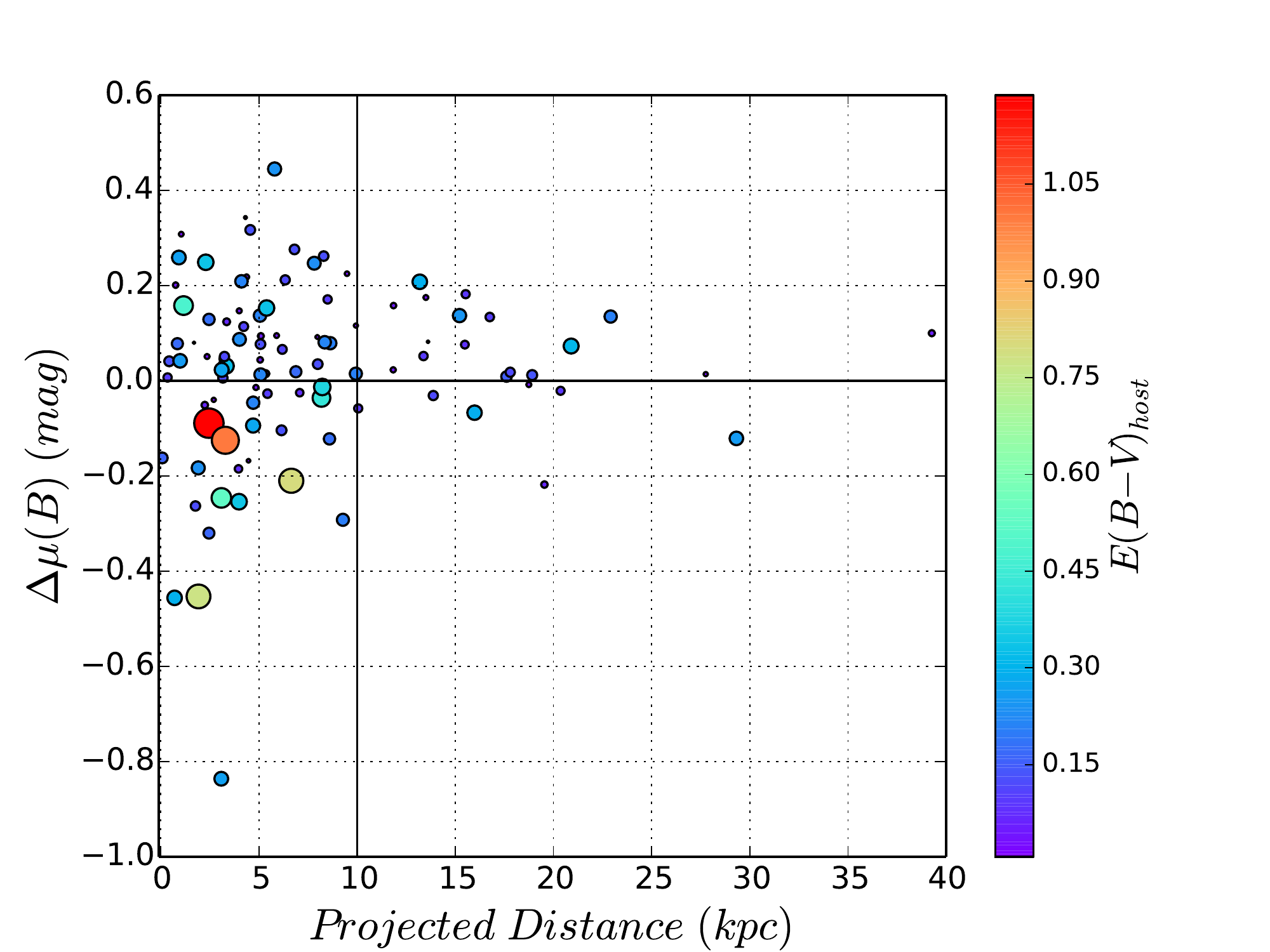} 
\caption{Plot of SN~Ia $B$-band Hubble residual vs. projected distances from host centers. Symbols in this dispersion plot are varying according to respective color excess values, $E(B-V)_{\rm host}$. Redder and larger symbols refer to higher color excess. SNe~Ia with larger color excess tend to reside closer to hosts and are more luminous.}
\label{ebv}
\end{center}
\end{figure}

\begin{figure*}[htbp]
\begin{center}
\begin{tabular}{cc}
\includegraphics[width=\columnwidth]{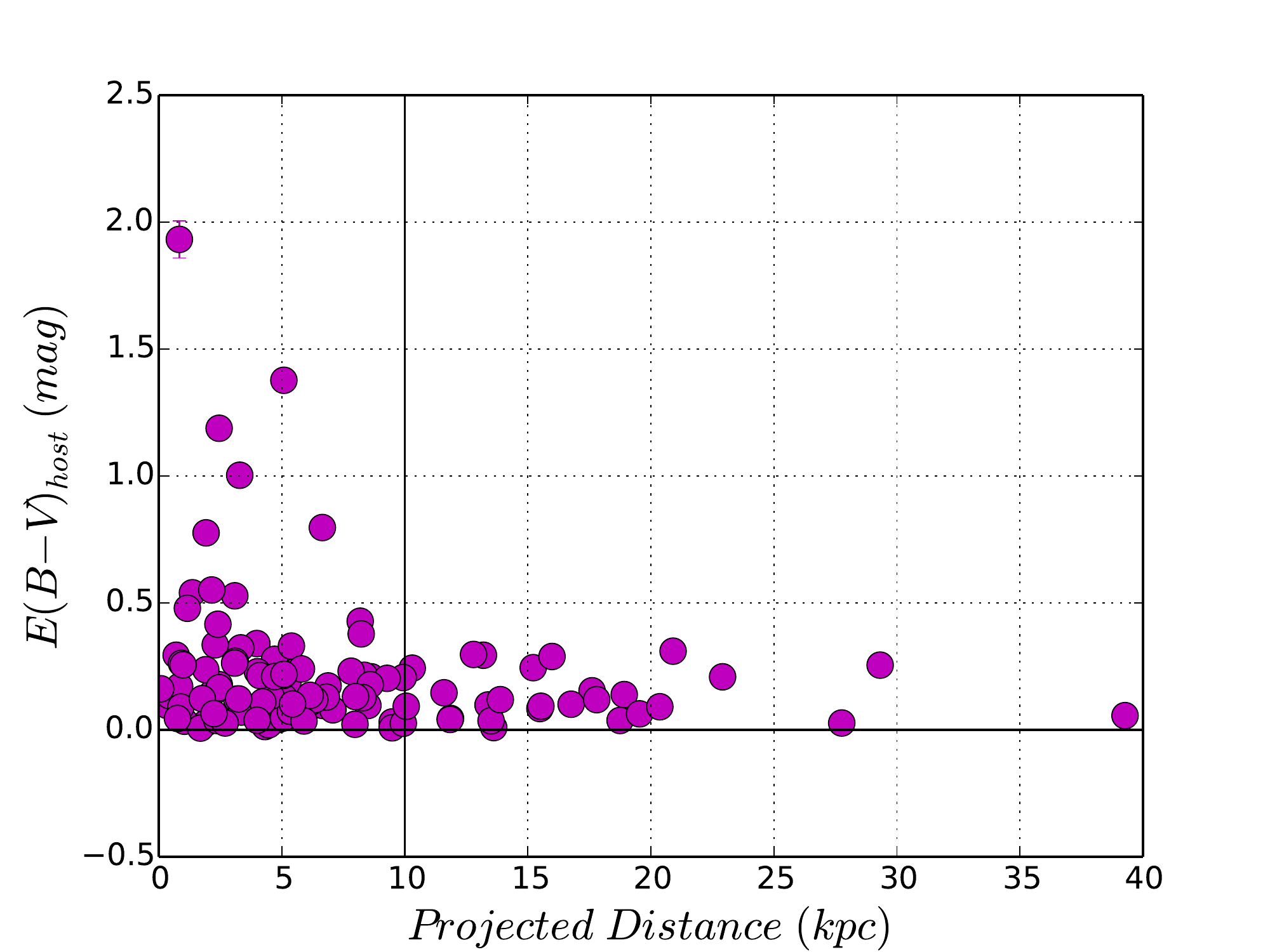} &
\includegraphics[width=\columnwidth]{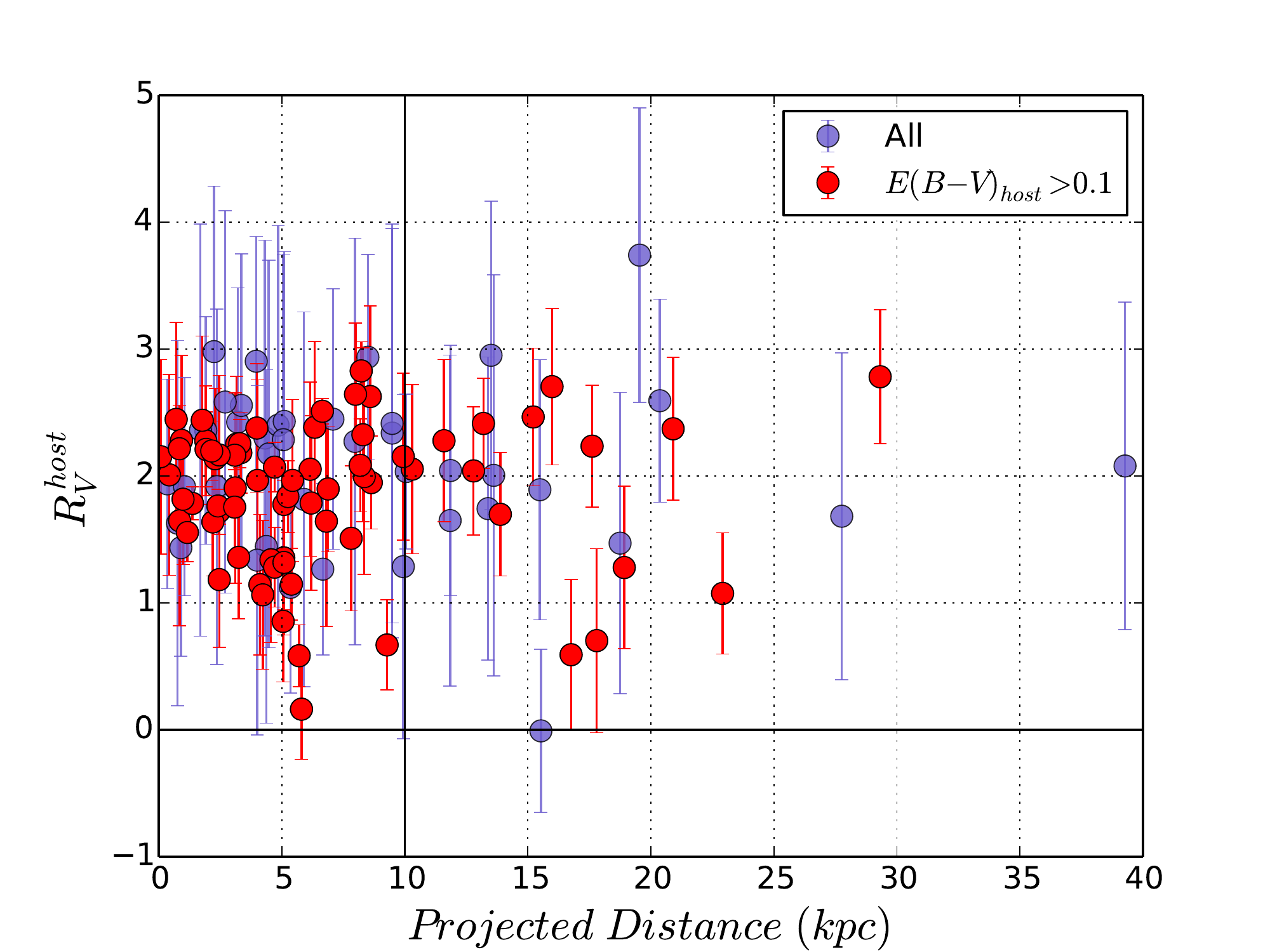} 
\end{tabular}
\caption{\emph{Left:} Variation of color excess $E(B-V)_{\rm host}$, and extinction law $R_V^{\rm host}$ , at the position of SN~Ia as a function of the projected distance. On average, SNe~Ia exploding beyond 10 kpc have $E(B-V)_{host}<0.5$ mag. The largest color excess is observed for SN~2008cd, which is within 5.5 arcseconds from its host NGC 5038. \emph{Right:} Variation of extinction law $R_V^{\rm host}$  at the position of SNe~Ia as a function of the projected distance. We do not observe significant variation of $R_V^{\rm host}$ .}
\label{rv}
\end{center}
\end{figure*}

 %Discussion (c): When you say "Contrary to Hicken+09, we do not find a significant difference in the weighted means ...", that doesn't sound quite right. You could find a significant result or a not significant result and that by itself is not consistent or inconsistent with Hicken+09. What matters is whether your measured difference is consistent or inconsistent with Hicken+09. Hicken+09 found that SNeIa in Scd/Sd/Irr hosts are 0.144 +- 0.07 mag fainter than those in E/S0. You should clearly state your quantitative result here. I interpret Table~4 of "0.002 +- 0.09" mag. The one-sigma errorbars of each measurement noticeably overlap.

\item \citet{hicken09} found that SNe~Ia in spiral galaxies (Scd/Sd/Irr) are intrinsically fainter than their E/S0 counterparts by $0.14 \pm 0.07$ mag, after light-curve correction. We have used galaxy morphological information in \autoref{sec:mor}, and use it again to examine how SN~Ia Hubble residuals are distributed when they are grouped according to host morphological types. We do not find a significant difference in the weighted means of $\Delta \mu (B)$ between spiral hosts and E/S0 hosts as we measure a difference of $0.02 \pm 0.09$ mag in $\Delta \mu (B)$ between them. We summarize the weighted means in Table~\ref{mort}.  A recent study by \citet{maria20} also supports our findings where they did not observe any difference in $\Delta \mu (B)$ between Early-type and Late-type SN~Ia host galaxies.

%I totally agree that morphological classification is "not homogeneous" to use kind words. But morphological classification is a rich field with a long history. The reference to Barchi+20 seems an almost out-of-place throw-away sentence. I don't think the fundamental issue is one of pattern recognition. I think it's that we have no idea if morphology actually means anything fundamental nor how to consistently measure morphology across different physical resolution and absolute surface brightness limits with redshift.

We note that galaxy morphology from NED is not homogeneous. The morphological classification method and images that are used to classify galaxies are also variable. A consistent classification scheme is therefore preferable. 

%One such avenue will be a machine learning based galaxy morphological analysis such as the one described in \citet{barchi20}. 

%\begin{figure}[htbp]
%\begin{center}
%\includegraphics[width=\columnwidth]{CSPI_resmorph.pdf} 
%\caption{Distribution of SN~Ia Hubble residual in various morphological types. Means of different distributions are consistent to each other.}
%\label{resmor}
%\end{center}
%\end{figure}
%

%* Right justify numbers in columns so that they line up more readably. E.g., "No." and "Weighted mean..." in Table~4 right-justified would line up and be more readable. But even more so for the tables in the appendix. E.g., ever negative sign moves the numbers over so they don't line up. It's somewhat distracting when trying to quickly scan through and get a sense of the numbers.

\begin{table}[htp]
\caption{Weighted mean of SN~Ia Hubble residual in different host morphological types.} 
\begin{center}
\begin{tabular}{lrr}
\hline
Morphology & No. & Weighted mean of $\Delta \mu (B)$\\
&&$(mag)$\\
\hline
\hline
%All &126 &$0.029 \pm 0.037$\\
Spiral & 59&$0.026 \pm 0.051$\\
E & 9&$-0.038 \pm 0.137$\\
S0 & 21&$0.056 \pm 0.087$\\
E +S0 &30 & $0.028 \pm 0.074$\\
\hline
\end{tabular}
\end{center}
\label{mort}
\end{table}%

\end{enumerate}

%\section{Discussion}\label{sec:dis}

\section{Conclusion}\label{sec:con}
We have presented photometry of host galaxies of SNe~Ia from the CSP-I in $ugriYJH$ filters. We have determined host galaxy stellar masses, and have studied the correlation between host galaxy stellar mass and SNe~Ia Hubble residual in the $uBgVriYJH$ bands. We emphasize that this is the first time such a study has been made using optical and near-infrared Hubble residuals of SNe~Ia. Our results are consistent with previous studies that SNe~Ia are on average more luminous in massive hosts, after correcting for the luminosity-decline rate relation. Moreover, we find that the Hubble residual offset, $\Delta_{\rm HR}$, is approximately constant from optical to near-infrared wavelengths, suggesting that dust in SNe~Ia host galaxies does not play a large role in driving these correlations.  We have also determined projected distances between the SNe and the centers of their host galaxies.  We find that SNe~Ia that explode beyond 10 kpc from their host galaxy centers have a smaller dispersion in their luminosities compared to those that explode within 10 kpc.  These SNe also have a smaller variation in color excess than do SNe that explode within 10 kpc.  SNe~Ia exploding closer to host centers also more commonly have higher color excesses.  These results should be confirmed with a larger sample with host galaxies clearly identified. We did not find differences in SN~Ia standardized luminosities when they are grouped according to their host morphological types.

We are gathering integral-field spectroscopy of CSP SN~Ia host galaxies (e.g., \citealt{galbany18}) to study their local environments. This may provide more insight into the origin of the observed luminosity offsets along with other findings.

\acknowledgments

We thank the anonymous referee for helping to improve this paper. The work of the $Carnegie \ Supernova \ Project$ has been supported by the National Science Foundation under grants AST0306969, AST0607438, AST1008343, AST1613426, AST1613472, and AST613455. NBS and KK gratefully acknowledge the support from the George P. and Cynthia Woods Mitchell Institute for Fundamental Physics and Astronomy. We also thank the Mitchell Foundation for their support of the Cooks Branch Workshop on Supernovae. M. S. is funded by generous grants from the Villum FONDEN No. 24349 and the Independent Research Fund Denmark No. 29774. L.G. was funded by the European Union's Horizon 2020 research and innovation program under the Marie Sk\l{}odowska-Curie grant agreement No. 839090. This work has been partially supported by the Spanish grant PGC2018-095317-B-C21 within the European Funds for Regional Development (FEDER).

\appendix

\section{SN~Ia Hubble Residual}
We include multi-band Hubble residual for each SN Ia used in this work.
\setcounter{table}{0}
\renewcommand{\thetable}{\Alph{section}\arabic{table}}

\begin{longrotatetable}
\begin{deluxetable*}{llccccccccc}
\tablecaption{CSP-I SN~Ia Hubble Residuals\label{restable}}
%\tablewidth{500pt}
\tabletypesize{\scriptsize}
\tablehead{
\colhead{$SN  \ Name$}  & \colhead{$z_{CMB}$} & \colhead{$u$} &
\colhead{$B$} &
\colhead{$g$} & \colhead{$V$}&  \colhead{$r$} & 
\colhead{$i$} & \colhead{$Y$} & 
\colhead{$J$} & \colhead{$H$}
} 
\startdata
2004ef&0.031&	 0.13(0.20)	& 0.02(0.15)	& 0.02(0.12)	&$-$0.00(0.15)	& 0.01(0.15)	& 0.07(0.16)	&$-$0.00(0.20)	& 0.08(0.24)	&$-$0.05(0.20)	\\
2004eo&0.016&	$-$0.13(0.21)	& 0.01(0.14)	& 0.00(0.11)	& 0.04(0.14)	&$-$0.03(0.15)	&$-$0.05(0.16)	& 0.03(0.18)	& 0.02(0.21)	& 0.09(0.19)	\\
2004ey&0.016&	 0.15(0.23)	& 0.22(0.14)	& 0.19(0.11)	& 0.22(0.14)	& 0.19(0.14)	& 0.16(0.15)	& 0.24(0.18)	& 0.17(0.21)	& 0.34(0.19)	\\
2004gs&0.027&	 0.11(0.20)	& 0.08(0.14)	& 0.09(0.11)	& 0.09(0.14)	& 0.09(0.14)	& 0.12(0.15)	& 0.16(0.19)	& 0.07(0.23)	& 0.23(0.20)	\\
2004gu&0.046&	 0.05(0.22)	& 0.13(0.16)	& 0.12(0.13)	& 0.07(0.16)	& 0.12(0.16)	& 0.13(0.16)	& 0.13(0.19)	& 0.12(0.29)	& 0.29(0.22)	\\
2005A&0.019&	$\cdots$&$-$0.09(0.16)	&$-$0.10(0.14)	&$-$0.07(0.16)	& 0.03(0.16)	& 0.10(0.16)	&$-$0.20(0.21)	&$-$0.05(0.24)	&$-$0.04(0.21)	\\
2005ag&0.079&	$\cdots$& 0.05(0.17)	& 0.07(0.14)	& 0.08(0.17)	& 0.08(0.17)	& 0.04(0.17)	& 0.13(0.20)	& 0.22(0.24)	&$\cdots$\\
2005al&0.012&	 0.40(0.20)	& 0.34(0.13)	& 0.33(0.09)	& 0.35(0.13)	& 0.34(0.13)	& 0.37(0.14)	& 0.35(0.19)	& 0.28(0.22)	& 0.36(0.20)	\\
2005am&0.008&	$-$0.48(0.20)	&$-$0.40(0.14)	&$-$0.39(0.11)	&$-$0.39(0.14)	&$-$0.41(0.14)	&$-$0.35(0.15)	&$-$0.35(0.18)	&$-$0.49(0.22)	&$-$0.30(0.19)	\\
2005be&0.035&	$-$0.09(0.24)	&$-$0.01(0.17)	&$-$0.03(0.14)	&$-$0.04(0.16)	&$-$0.01(0.15)	&$-$0.01(0.16)	&$\cdots$&$\cdots$&$\cdots$\\
2005bg&0.023&	$-$0.04(0.23)	& 0.01(0.17)	&$-$0.00(0.14)	&$-$0.03(0.16)	&$-$0.00(0.16)	& 0.02(0.17)	&$\cdots$&$\cdots$&$\cdots$\\
2005bl&0.024&	 0.00(0.27)	& 0.08(0.22)	& 0.04(0.20)	& 0.04(0.21)	&$-$0.04(0.22)	& 0.07(0.20)	&$\cdots$&$\cdots$&$\cdots$\\
2005bo&0.014&	$-$0.23(0.23)	&$-$0.25(0.20)	&$-$0.28(0.18)	&$-$0.23(0.20)	&$-$0.26(0.20)	&$-$0.26(0.19)	&$\cdots$&$\cdots$&$\cdots$\\
2005el&0.015&	$-$0.05(0.19)	& 0.08(0.13)	& 0.05(0.09)	& 0.09(0.13)	& 0.11(0.13)	& 0.06(0.14)	& 0.09(0.17)	& 0.12(0.21)	& 0.14(0.19)	\\
2005eq&0.029&	$-$0.10(0.20)	& 0.01(0.14)	&$-$0.03(0.11)	& 0.01(0.14)	& 0.03(0.14)	& 0.03(0.15)	& 0.06(0.18)	&$-$0.00(0.21)	& 0.10(0.19)	\\
2005hc&0.046&	 0.25(0.20)	& 0.17(0.14)	& 0.15(0.11)	& 0.16(0.14)	& 0.15(0.14)	& 0.18(0.15)	& 0.15(0.18)	& 0.27(0.21)	& 0.06(0.21)	\\
2005hj&0.058&	$-$0.13(0.25)	& 0.05(0.18)	& 0.03(0.15)	&$-$0.00(0.17)	&$-$0.00(0.17)	& 0.06(0.17)	&$-$0.02(0.20)	& 0.02(0.25)	&$\cdots$\\
2005iq&0.034&	 0.09(0.20)	& 0.17(0.13)	& 0.14(0.10)	& 0.15(0.13)	& 0.18(0.13)	& 0.14(0.14)	& 0.14(0.18)	& 0.14(0.21)	& 0.21(0.19)	\\
2005ir&0.076&	$-$0.02(0.21)	& 0.04(0.16)	& 0.01(0.14)	&$-$0.01(0.16)	& 0.07(0.16)	& 0.02(0.16)	&$\cdots$&$\cdots$&$\cdots$\\
2005kc&0.015&	$-$0.06(0.20)	& 0.03(0.14)	& 0.02(0.11)	& 0.02(0.14)	& 0.03(0.14)	&$-$0.04(0.15)	&$-$0.02(0.18)	& 0.00(0.21)	& 0.10(0.19)	\\
2005ke&0.005&	 0.45(0.21)	& 0.66(0.14)	& 0.60(0.11)	& 0.61(0.15)	& 0.61(0.14)	& 0.68(0.15)	& 0.60(0.18)	& 0.71(0.21)	& 0.56(0.19)	\\
2005ki&0.019&	$-$0.14(0.20)	& 0.01(0.13)	&$-$0.01(0.09)	& 0.01(0.13)	& 0.01(0.13)	&$-$0.02(0.15)	& 0.03(0.18)	& 0.04(0.21)	& 0.06(0.19)	\\
2005ku&0.045&	 0.32(0.26)	& 0.08(0.20)	& 0.12(0.18)	& 0.06(0.20)	& 0.06(0.19)	& 0.09(0.19)	& 0.01(0.27)	&$\cdots$&$\cdots$\\
2005lu&0.032&	 0.19(0.25)	& 0.25(0.21)	& 0.17(0.18)	& 0.23(0.21)	& 0.24(0.21)	& 0.24(0.20)	& 0.28(0.21)	&$\cdots$&$\cdots$\\
2005M&0.022&	$-$0.09(0.20)	& 0.01(0.14)	&$-$0.01(0.11)	& 0.02(0.14)	&$-$0.00(0.14)	&$-$0.10(0.15)	& 0.01(0.18)	& 0.22(0.21)	&$-$0.02(0.19)	\\
2005mc&0.025&	 0.03(0.28)	& 0.02(0.25)	& 0.02(0.23)	& 0.02(0.24)	& 0.01(0.23)	& 0.04(0.22)	&$\cdots$&$\cdots$&$\cdots$\\
2005na&0.026&	$-$0.49(0.20)	&$-$0.18(0.14)	&$-$0.21(0.11)	&$-$0.20(0.14)	&$-$0.20(0.14)	&$-$0.23(0.15)	&$-$0.07(0.18)	&$-$0.23(0.24)	&$-$0.14(0.19)	\\
2005W&0.009&	 0.19(0.22)	& 0.18(0.20)	& 0.16(0.17)	& 0.17(0.20)	& 0.18(0.20)	& 0.18(0.19)	&$\cdots$&$\cdots$&$\cdots$\\
2006ax&0.017&	$-$0.09(0.20)	&$-$0.01(0.13)	&$-$0.04(0.10)	&$-$0.02(0.13)	&$-$0.02(0.13)	&$-$0.07(0.14)	& 0.03(0.18)	&$-$0.07(0.21)	& 0.07(0.19)	\\
2006bd&0.026&	$\cdots$& 0.25(0.18)	& 0.19(0.16)	& 0.15(0.18)	& 0.18(0.18)	& 0.15(0.19)	& 0.05(0.20)	& 0.18(0.24)	& 0.36(0.23)	\\
2006bh&0.011&	 0.06(0.19)	& 0.16(0.13)	& 0.14(0.10)	& 0.17(0.13)	& 0.15(0.13)	& 0.16(0.14)	& 0.20(0.18)	& 0.18(0.21)	& 0.18(0.19)	\\
2006br&0.025&	$\cdots$&$-$0.12(0.18)	&$-$0.24(0.15)	&$-$0.20(0.19)	&$-$0.10(0.19)	&$-$0.04(0.19)	&$-$0.14(0.20)	&$-$0.27(0.24)	&$-$0.01(0.22)	\\
2006D&0.009&	$-$0.01(0.19)	&$-$0.06(0.13)	&$-$0.10(0.10)	&$-$0.07(0.14)	&$-$0.06(0.14)	&$-$0.07(0.15)	& 0.06(0.18)	&$-$0.01(0.22)	&$-$0.10(0.19)	\\
2006ef&0.018&	 0.20(0.25)	& 0.23(0.17)	& 0.21(0.13)	& 0.20(0.16)	& 0.23(0.15)	& 0.23(0.16)	&$\cdots$&$\cdots$&$\cdots$\\
2006ej&0.021&	 0.09(0.23)	& 0.12(0.16)	& 0.11(0.12)	& 0.10(0.15)	& 0.10(0.15)	& 0.16(0.16)	& 0.08(0.18)	&$-$0.01(0.22)	& 0.18(0.20)	\\
2006eq&0.050&	 0.17(0.30)	& 0.01(0.18)	& 0.01(0.15)	&$-$0.01(0.17)	& 0.01(0.17)	& 0.07(0.18)	& 0.14(0.19)	& 0.12(0.24)	&$-$0.36(0.25)	\\
2006et&0.022&	$-$0.04(0.19)	&$-$0.09(0.13)	&$-$0.12(0.10)	&$-$0.08(0.13)	&$-$0.09(0.13)	&$-$0.14(0.14)	&$-$0.09(0.17)	&$-$0.04(0.21)	&$-$0.02(0.19)	\\
2006ev&0.029&	$\cdots$& 0.14(0.16)	& 0.14(0.13)	& 0.14(0.16)	& 0.17(0.16)	& 0.22(0.17)	& 0.21(0.19)	& 0.16(0.22)	& 0.15(0.20)	\\
2006fw&0.083&	$-$0.57(0.32)	&$-$0.46(0.27)	&$-$0.58(0.25)	&$-$0.34(0.27)	&$-$0.51(0.26)	&$-$0.55(0.24)	&$\cdots$&$\cdots$&$\cdots$\\
2006gj&0.028&	 0.15(0.21)	& 0.21(0.15)	& 0.22(0.12)	& 0.22(0.15)	& 0.20(0.15)	& 0.19(0.16)	& 0.18(0.19)	& 0.14(0.23)	& 0.35(0.21)	\\
2006gt&0.045&	 0.04(0.22)	& 0.08(0.15)	& 0.06(0.12)	& 0.04(0.15)	& 0.12(0.15)	& 0.08(0.16)	& 0.16(0.18)	& 0.18(0.22)	& 0.27(0.23)	\\
2006hb&0.015&	 0.13(0.22)	& 0.07(0.15)	& 0.04(0.11)	& 0.04(0.15)	& 0.09(0.15)	& 0.06(0.16)	& 0.08(0.19)	& 0.04(0.22)	& 0.13(0.19)	\\
2006hx&0.045&	$-$0.14(0.21)	&$-$0.18(0.15)	&$-$0.17(0.12)	&$-$0.19(0.15)	&$-$0.13(0.15)	&$-$0.31(0.16)	&$-$0.03(0.19)	&$-$0.13(0.22)	&$-$0.14(0.20)	\\
2006is&0.031&	$-$0.08(0.29)	&$-$0.17(0.18)	&$-$0.16(0.16)	&$-$0.17(0.17)	&$-$0.07(0.17)	& 0.11(0.19)	& 0.28(0.22)	& 0.02(0.28)	& 0.34(0.26)	\\
2006kf&0.021&	 0.01(0.21)	& 0.09(0.15)	& 0.09(0.12)	& 0.11(0.15)	& 0.08(0.14)	& 0.10(0.15)	& 0.09(0.18)	& 0.10(0.21)	& 0.10(0.20)	\\
2006lu&0.053&	$-$0.11(0.25)	& 0.04(0.17)	&$-$0.01(0.14)	& 0.01(0.16)	& 0.03(0.16)	&$-$0.01(0.17)	& 0.28(0.19)	&$\cdots$&$\cdots$\\
2006mr&0.006&	$-$0.86(0.35)	&$-$0.69(0.24)	&$-$0.60(0.22)	&$-$0.38(0.23)	&$-$0.34(0.23)	&$-$0.50(0.24)	&$-$0.43(0.26)	&$-$0.30(0.30)	&$-$0.37(0.28)	\\
2006ob&0.059&	$-$0.46(0.21)	&$-$0.03(0.15)	&$-$0.04(0.12)	&$-$0.06(0.15)	& 0.01(0.15)	&$-$0.07(0.15)	& 0.09(0.20)	& 0.12(0.24)	& 0.04(0.25)	\\
2006os&0.033&	$-$0.02(0.25)	&$-$0.04(0.17)	&$-$0.06(0.14)	&$-$0.05(0.17)	&$-$0.08(0.17)	&$-$0.10(0.18)	&$-$0.07(0.19)	&$-$0.03(0.21)	&$-$0.04(0.22)	\\
2006py&0.058&	$-$0.29(0.22)	&$-$0.16(0.18)	&$-$0.22(0.15)	&$-$0.20(0.17)	&$-$0.15(0.17)	&$-$0.16(0.17)	&$\cdots$&$\cdots$&$\cdots$\\
2006X&0.005&	$-$0.52(0.19)	&$-$1.16(0.13)	&$-$1.23(0.10)	&$-$1.22(0.14)	&$-$1.11(0.14)	&$-$1.00(0.15)	&$-$1.21(0.18)	&$-$1.26(0.21)	&$-$1.11(0.19)	\\
2007af&0.005&	 0.25(0.19)	& 0.25(0.13)	& 0.19(0.10)	& 0.26(0.13)	& 0.24(0.13)	& 0.23(0.14)	& 0.25(0.17)	& 0.26(0.21)	& 0.27(0.19)	\\
2007ai&0.032&	$-$0.13(0.25)	&$-$0.07(0.17)	&$-$0.06(0.15)	&$-$0.07(0.17)	&$-$0.05(0.17)	&$-$0.10(0.17)	&$-$0.07(0.18)	&$-$0.09(0.22)	&$-$0.18(0.19)	\\
2007al&0.012&	 0.25(0.28)	& 0.31(0.19)	& 0.25(0.16)	& 0.28(0.18)	& 0.24(0.17)	& 0.18(0.19)	& 0.23(0.21)	& 0.19(0.25)	&$-$0.01(0.22)	\\
2007as&0.018&	 0.20(0.21)	& 0.08(0.14)	& 0.06(0.11)	& 0.04(0.14)	& 0.03(0.14)	& 0.11(0.15)	& 0.08(0.18)	& 0.04(0.21)	& 0.12(0.19)	\\
2007ax&0.007&	 0.93(0.27)	& 0.49(0.17)	& 0.47(0.15)	& 0.45(0.17)	& 0.41(0.17)	& 0.40(0.17)	& 0.34(0.19)	&$-$0.56(0.25)	& 0.35(0.23)	\\
2007ba&0.038&	$-$0.61(0.21)	&$-$0.29(0.14)	&$-$0.33(0.11)	&$-$0.28(0.14)	&$-$0.32(0.14)	&$-$0.32(0.15)	&$-$0.18(0.18)	&$-$0.13(0.22)	&$-$0.32(0.21)	\\
2007bc&0.021&	$-$0.10(0.19)	&$-$0.03(0.14)	&$-$0.03(0.10)	&$-$0.03(0.14)	&$-$0.04(0.14)	&$-$0.07(0.15)	&$-$0.06(0.18)	& 0.06(0.21)	& 0.01(0.20)	\\
2007bd&0.031&	$-$0.06(0.21)	& 0.01(0.13)	&$-$0.03(0.10)	& 0.01(0.14)	&$-$0.01(0.14)	& 0.01(0.15)	& 0.04(0.18)	& 0.04(0.22)	& 0.06(0.20)	\\
2007bm&0.006&	$-$0.15(0.20)	&$-$0.25(0.13)	&$-$0.28(0.10)	&$-$0.24(0.14)	&$-$0.25(0.14)	&$-$0.32(0.15)	&$-$0.22(0.18)	&$-$0.25(0.21)	&$-$0.22(0.19)	\\
2007ca&0.014&	 0.05(0.20)	&$-$0.01(0.13)	&$-$0.04(0.10)	&$-$0.00(0.14)	& 0.01(0.14)	&$-$0.02(0.15)	& 0.08(0.18)	& 0.06(0.21)	&$-$0.00(0.19)	\\
2007cg&0.033&	$-$0.26(0.31)	&$-$0.45(0.23)	&$-$0.59(0.20)	&$-$0.49(0.23)	&$-$0.45(0.23)	&$-$0.40(0.22)	&$-$0.39(0.21)	&$\cdots$&$\cdots$\\
2007hj&0.014&	 0.30(0.21)	& 0.32(0.15)	& 0.32(0.11)	& 0.32(0.15)	& 0.30(0.15)	& 0.33(0.16)	& 0.33(0.19)	& 0.31(0.22)	& 0.40(0.20)	\\
2007hx&0.080&	 0.00(0.29)	& 0.07(0.21)	&$-$0.01(0.18)	& 0.04(0.21)	& 0.13(0.20)	& 0.02(0.20)	& 0.02(0.20)	&$\cdots$&$\cdots$\\
2007jd&0.073&	 0.74(0.32)	& 0.28(0.18)	& 0.29(0.14)	& 0.23(0.17)	& 0.30(0.17)	& 0.19(0.18)	& 0.29(0.21)	& 0.14(0.25)	&$\cdots$\\
2007jg&0.037&	 0.58(0.19)	& 0.21(0.14)	& 0.20(0.11)	& 0.20(0.14)	& 0.23(0.14)	& 0.26(0.15)	& 0.25(0.18)	& 0.20(0.21)	& 0.03(0.20)	\\
2007jh&0.041&	$\cdots$& 0.09(0.22)	& 0.07(0.20)	& 0.07(0.21)	& 0.08(0.21)	& 0.15(0.20)	&$-$0.07(0.30)	&$\cdots$&$\cdots$\\
2007le&0.007&	 0.57(0.19)	& 0.44(0.13)	& 0.40(0.10)	& 0.42(0.13)	& 0.46(0.13)	& 0.51(0.14)	& 0.49(0.18)	& 0.43(0.21)	& 0.49(0.19)	\\
2007N&0.013&	 1.00(0.33)	& 0.21(0.21)	& 0.09(0.18)	& 0.19(0.21)	& 0.10(0.21)	& 0.16(0.21)	& 0.03(0.23)	& 0.08(0.26)	&$-$0.01(0.23)	\\
2007nq&0.045&	$-$0.26(0.20)	&$-$0.22(0.14)	&$-$0.24(0.11)	&$-$0.23(0.14)	&$-$0.19(0.14)	&$-$0.19(0.15)	&$-$0.14(0.19)	&$-$0.23(0.23)	&$-$0.34(0.22)	\\
2007ol&0.056&	 0.05(0.20)	& 0.08(0.13)	& 0.06(0.10)	& 0.16(0.13)	& 0.14(0.13)	&$\cdots$&$\cdots$&$\cdots$&$\cdots$\\
2007on&0.006&	 0.02(0.20)	& 0.13(0.13)	& 0.14(0.10)	& 0.14(0.13)	& 0.12(0.13)	& 0.12(0.14)	& 0.16(0.18)	& 0.15(0.21)	& 0.14(0.19)	\\
2007S&0.014&	$-$0.31(0.19)	&$-$0.25(0.13)	&$-$0.31(0.10)	&$-$0.25(0.13)	&$-$0.24(0.13)	&$-$0.25(0.14)	&$-$0.22(0.17)	&$-$0.20(0.21)	&$-$0.20(0.19)	\\
2007st&0.021&	$-$0.41(0.23)	&$-$0.32(0.16)	&$-$0.31(0.13)	&$-$0.33(0.16)	&$-$0.32(0.16)	&$-$0.38(0.17)	&$-$0.34(0.19)	&$-$0.60(0.24)	&$-$0.14(0.21)	\\
2007ux&0.031&	 0.05(0.21)	& 0.11(0.14)	& 0.10(0.11)	& 0.10(0.14)	& 0.13(0.14)	& 0.14(0.15)	& 0.21(0.18)	& 0.20(0.21)	& 0.23(0.19)	\\
2008ar&0.026&	 0.07(0.19)	& 0.05(0.13)	& 0.02(0.10)	& 0.05(0.14)	& 0.02(0.14)	& 0.03(0.15)	& 0.12(0.18)	& 0.11(0.21)	& 0.00(0.19)	\\
2008bc&0.015&	 0.10(0.24)	& 0.10(0.14)	& 0.06(0.11)	& 0.07(0.14)	& 0.06(0.14)	& 0.07(0.15)	& 0.17(0.18)	& 0.12(0.21)	& 0.12(0.19)	\\
2008bi&0.013&	 0.44(0.31)	& 0.26(0.23)	& 0.16(0.19)	& 0.24(0.21)	& 0.36(0.20)	& 0.23(0.20)	& 0.33(0.19)	&$-$0.28(0.25)	& 0.35(0.21)	\\
2008bq&0.034&	 0.04(0.21)	&$-$0.12(0.16)	&$-$0.15(0.13)	&$-$0.09(0.16)	&$-$0.13(0.16)	&$-$0.11(0.16)	&$-$0.16(0.19)	&$-$0.10(0.24)	&$-$0.07(0.23)	\\
2008bt&0.015&	 0.02(0.22)	& 0.14(0.16)	& 0.11(0.13)	& 0.11(0.16)	& 0.10(0.16)	& 0.11(0.17)	& 0.23(0.19)	& 0.09(0.26)	& 0.26(0.31)	\\
2008bz&0.060&	 0.11(0.23)	& 0.09(0.15)	& 0.06(0.12)	& 0.07(0.14)	& 0.12(0.14)	& 0.12(0.16)	& 0.17(0.18)	& 0.20(0.23)	&$\cdots$\\
2008C&0.017&	 0.05(0.24)	& 0.04(0.17)	&$-$0.01(0.13)	& 0.02(0.17)	& 0.02(0.17)	&$-$0.04(0.18)	& 0.03(0.19)	&$-$0.05(0.23)	& 0.11(0.20)	\\
2008cc&0.010&	$-$0.20(0.22)	&$-$0.26(0.15)	&$-$0.25(0.11)	&$-$0.27(0.15)	&$-$0.23(0.14)	&$-$0.19(0.16)	&$-$0.19(0.18)	&$-$0.40(0.22)	&$-$0.23(0.19)	\\
2008cd&0.007&	$\cdots$&$-$0.46(0.59)	&$-$0.67(0.58)	&$-$0.47(0.60)	&$-$0.43(0.59)	&$-$0.45(0.49)	&$\cdots$&$\cdots$&$\cdots$\\
2008cf&0.046&	$-$0.11(0.23)	& 0.05(0.15)	&$-$0.01(0.12)	&$-$0.01(0.15)	& 0.02(0.14)	& 0.04(0.16)	& 0.29(0.19)	& 0.21(0.27)	& 0.00(0.20)	\\
2008fl&0.020&	$-$0.15(0.23)	&$-$0.10(0.16)	&$-$0.12(0.12)	&$-$0.10(0.15)	&$-$0.09(0.15)	&$-$0.17(0.17)	&$-$0.14(0.18)	&$-$0.12(0.22)	&$-$0.06(0.20)	\\
2008fp&0.006&	$-$0.33(0.20)	&$-$0.33(0.14)	&$-$0.38(0.11)	&$-$0.37(0.15)	&$-$0.29(0.15)	&$-$0.42(0.16)	&$-$0.31(0.18)	&$-$0.35(0.21)	&$-$0.25(0.19)	\\
2008fr&0.039&	 0.18(0.26)	& 0.20(0.15)	& 0.16(0.12)	& 0.16(0.15)	& 0.19(0.15)	& 0.18(0.16)	& 0.30(0.19)	& 0.39(0.23)	& 0.19(0.19)	\\
2008fu&0.052&	$-$0.06(0.22)	&$-$0.04(0.16)	&$-$0.00(0.12)	& 0.01(0.15)	& 0.03(0.14)	& 0.01(0.16)	&$-$0.03(0.19)	&$-$0.22(0.24)	& 0.01(0.20)	\\
2008fw&0.009&	$-$0.04(0.21)	& 0.14(0.15)	& 0.09(0.11)	& 0.13(0.15)	& 0.14(0.15)	& 0.12(0.16)	& 0.14(0.18)	& 0.28(0.21)	& 0.13(0.19)	\\
2008gg&0.032&	 0.14(0.24)	& 0.14(0.15)	& 0.09(0.12)	& 0.10(0.16)	& 0.10(0.16)	& 0.14(0.17)	& 0.20(0.18)	& 0.01(0.23)	& 0.27(0.19)	\\
2008gl&0.034&	 0.07(0.21)	& 0.13(0.14)	& 0.12(0.10)	& 0.13(0.14)	& 0.12(0.14)	& 0.12(0.15)	& 0.31(0.18)	& 0.02(0.21)	& 0.12(0.20)	\\
2008go&0.062&	 0.18(0.22)	& 0.02(0.15)	&$-$0.01(0.12)	& 0.01(0.15)	&$-$0.00(0.15)	& 0.07(0.16)	& 0.06(0.18)	&$\cdots$&$\cdots$\\
2008gp&0.033&	$-$0.16(0.22)	& 0.02(0.14)	&$-$0.02(0.10)	&$-$0.02(0.14)	& 0.00(0.14)	&$-$0.03(0.15)	& 0.02(0.18)	&$-$0.01(0.21)	& 0.16(0.20)	\\
2008hj&0.038&	 0.01(0.24)	& 0.18(0.13)	& 0.15(0.10)	& 0.18(0.14)	& 0.18(0.14)	& 0.19(0.15)	& 0.31(0.17)	& 0.24(0.21)	& 0.13(0.20)	\\
2008hu&0.050&	 0.17(0.22)	& 0.10(0.14)	& 0.10(0.11)	& 0.09(0.14)	& 0.09(0.14)	& 0.12(0.15)	& 0.15(0.19)	& 0.19(0.22)	& 0.00(0.22)	\\
2008hv&0.013&	 0.01(0.20)	& 0.12(0.13)	& 0.10(0.09)	& 0.12(0.13)	& 0.11(0.13)	& 0.11(0.14)	& 0.17(0.18)	& 0.08(0.21)	& 0.18(0.19)	\\
2008ia&0.022&	$-$0.01(0.21)	&$-$0.05(0.15)	&$-$0.05(0.12)	&$-$0.01(0.15)	&$-$0.04(0.15)	&$-$0.01(0.15)	& 0.02(0.18)	&$-$0.04(0.21)	&$-$0.16(0.19)	\\
2008O&0.039&	 0.34(0.21)	& 0.15(0.15)	& 0.16(0.12)	& 0.17(0.15)	& 0.14(0.15)	& 0.16(0.16)	& 0.16(0.18)	& 0.18(0.22)	& 0.17(0.20)	\\
2008R&0.013&	$-$0.02(0.24)	& 0.15(0.14)	& 0.14(0.11)	& 0.15(0.14)	& 0.11(0.14)	& 0.12(0.15)	& 0.22(0.19)	& 0.18(0.22)	& 0.25(0.20)	\\
2009aa&0.027&	$-$0.22(0.19)	&$-$0.06(0.13)	&$-$0.09(0.10)	&$-$0.07(0.13)	&$-$0.06(0.14)	&$-$0.12(0.14)	&$-$0.03(0.18)	&$-$0.07(0.21)	&$-$0.02(0.19)	\\
2009ab&0.011&	 0.08(0.20)	& 0.26(0.14)	& 0.25(0.11)	& 0.27(0.14)	& 0.27(0.14)	& 0.22(0.15)	& 0.20(0.18)	& 0.21(0.21)	& 0.32(0.19)	\\
2009ad&0.028&	$-$0.32(0.20)	&$-$0.03(0.13)	&$-$0.06(0.10)	&$-$0.05(0.14)	&$-$0.03(0.14)	&$-$0.08(0.15)	&$-$0.03(0.18)	&$-$0.07(0.21)	& 0.09(0.19)	\\
2009ag&0.009&	 0.37(0.21)	& 0.18(0.15)	& 0.16(0.13)	& 0.21(0.16)	& 0.21(0.16)	& 0.15(0.17)	& 0.14(0.18)	& 0.16(0.21)	& 0.22(0.19)	\\
2009al&0.022&	$-$0.32(0.24)	&$-$0.12(0.16)	&$-$0.17(0.13)	&$-$0.10(0.16)	&$-$0.05(0.16)	&$-$0.27(0.18)	&$-$0.20(0.18)	&$-$0.12(0.21)	&$-$0.05(0.20)	\\
2009cz&0.021&	 0.07(0.20)	& 0.04(0.13)	& 0.01(0.11)	& 0.04(0.14)	& 0.01(0.14)	&$-$0.04(0.15)	& 0.03(0.18)	& 0.06(0.21)	& 0.10(0.20)	\\
2009D&0.025&	 0.07(0.20)	&$-$0.02(0.14)	&$-$0.06(0.10)	&$-$0.00(0.14)	&$-$0.03(0.14)	&$-$0.08(0.14)	& 0.01(0.19)	&$-$0.01(0.21)	& 0.01(0.19)	\\
2009ds&0.019&	$-$0.16(0.20)	&$-$0.05(0.14)	&$-$0.07(0.10)	&$-$0.05(0.14)	&$-$0.06(0.14)	&$-$0.09(0.15)	& 0.21(0.19)	&$-$0.10(0.21)	& 0.05(0.19)	\\
2009F&0.013&	$-$1.38(0.25)	&$-$0.84(0.18)	&$-$0.71(0.15)	&$-$0.45(0.18)	&$-$0.48(0.18)	&$-$0.28(0.18)	&$-$0.20(0.21)	&$-$0.06(0.24)	&$-$0.03(0.22)	\\
2009I&0.026&	 0.05(0.20)	&$-$0.21(0.14)	&$-$0.23(0.10)	&$-$0.19(0.14)	&$-$0.20(0.14)	&$-$0.21(0.15)	&$-$0.22(0.18)	&$-$0.23(0.21)	&$-$0.11(0.19)	\\
2009le&0.018&	$-$0.04(0.23)	& 0.45(0.15)	& 0.42(0.13)	& 0.45(0.16)	& 0.42(0.16)	& 0.38(0.16)	&$\cdots$& 0.71(0.21)	&$\cdots$\\
2009P&0.025&	 0.21(0.20)	& 0.16(0.16)	& 0.10(0.13)	& 0.13(0.16)	& 0.16(0.16)	& 0.22(0.16)	& 0.27(0.19)	& 0.08(0.22)	&$\cdots$\\
2009Y&0.009&	 0.03(0.20)	& 0.01(0.14)	& 0.00(0.11)	&$-$0.02(0.14)	&$-$0.03(0.14)	& 0.01(0.15)	& 0.00(0.18)	& 0.13(0.21)	& 0.00(0.19)	\\
\enddata
\end{deluxetable*}
\end{longrotatetable}

\section {Host Galaxy Photometry}
We include a table of $ugriYJH$ photometry of CSP-I SN~Ia host galaxies, Values within parentheses are uncertainties of measurements. When photometry is not available, it is shown as $\cdots$. Morphological classifications are taken from the NED.

\setcounter{table}{0}

\begin{longrotatetable}
\begin{deluxetable*}{llcccccccc}
\tablecaption{CSP-I Host Galaxy Photometry\label{phottable}}
%\tablewidth{500pt}
\tabletypesize{\scriptsize}
\tablehead{
\colhead{$SN  \ Name$} & \colhead{$Host \ Galaxy \ Name$} &\colhead{Morphology} & \colhead{$u$} &
\colhead{$g$} & \colhead{$r$} & 
\colhead{$i$} & \colhead{$Y$} & 
\colhead{$J$} & \colhead{$H$}
} 
\startdata
2004ef& UGC 12158&SA&17.44(08)&14.07(04)&13.40(06)&13.12(03)&12.39(12)&11.77(12)&11.12(09)\\
2004eo& NGC 6928&SB&14.64(08)&12.68(05)&11.83(08)&11.09(13)&$\cdots$&9.81(02)&9.07(03)\\
2004ey& UGC 11816&SB&18.02(08)&13.90(06)&13.44(11)&13.18(11)&12.08(05)&12.23(06)&11.54(08)\\
2004gs& MCG +03-22-020&S&16.65(08)&14.65(06)&13.72(11)&13.32(11)&12.46(10)&12.05(09)&11.33(09)\\
2004gu& FGC 175A&$\cdots$&19.24(09)&17.23(08)&16.48(08)&16.01(08)&14.53(09)&14.07(10)&13.16(12)\\
2005A& NGC 958&SB&14.29(09)&12.49(11)&11.85(09)&11.41(12)&10.49(12)&9.91(14)&9.03(16)\\
2005ag& 2MASS J14564320+0919366&$\cdots$&$\cdots$&16.26(08)&15.49(07)&15.06(08)&14.14(07)&13.57(12)&12.97(10)\\
2005al& NGC 5304&cD&15.17(10)&13.22(07)&12.37(06)&11.99(07)&10.93(16)&10.14(14)&9.94(12)\\
2005am& NGC 2811&SB&13.55(09)&11.70(09)&10.93(09)&10.50(09)&9.32(09)&8.90(08)&8.38(08)\\
2005be& 2MASS J14593308+1640068&$\cdots$&17.67(07)&15.45(11)&14.51(11)&14.16(06)&$\cdots$&12.76(04)&12.01(05)\\
2005bg& MCG +03-31-93&$\cdots$&15.89(14)&14.79(12)&14.24(14)&14.02(13)&$\cdots$&12.96(04)&12.33(06)\\
2005bl& NGC 4059&E&15.55(14)&13.48(11)&12.68(09)&12.20(06)&$\cdots$&10.73(02)&10.01(02)\\
2005bo& NGC 4708&SA&16.03(07)&13.79(10)&13.10(10)&12.40(04)&$\cdots$&11.16(03)&10.52(04)\\
2005el& NGC 1819&S0&15.38(10)&13.37(06)&12.17(09)&11.79(07)&10.89(09)&10.44(10)&9.78(10)\\
2005eq& MCG -01-09-006&SB&17.00(09)&14.18(03)&13.53(08)&13.16(10)&12.75(14)&12.36(15)&11.75(16)\\
2005hc& MCG +00-06-003&$\cdots$&18.50(11)&15.66(08)&14.91(10)&14.54(10)&13.59(13)&13.21(20)&12.55(12)\\
2005hj& APMUKS(BJ) B012415.21-012950.3&$\cdots$&20.05(12)&18.99(02)&18.26(01)&17.88(02)&16.81(10)&16.24(11)&15.43(28)\\
2005iq& MCG -03-01-008&SA&17.76(09)&15.01(08)&14.29(09)&13.94(08)&13.08(12)&12.60(18)&12.07(19)\\
2005ir& SDSS J011643.87+004736.9&$\cdots$&18.58(04)&17.40(08)&16.87(07)&16.49(06)&$\cdots$&15.77(14)&15.01(19)\\
2005kc& NGC 7311&SA&14.25(15)&12.50(05)&11.76(07)&11.33(07)&10.40(09)&9.22(07)&8.67(09)\\
2005ke& NGC 1371&SA&14.85(13)&11.74(05)&10.84(09)&10.43(03)&$\cdots$&8.75(02)&8.08(02)\\
2005ki& NGC 3332&SA&15.33(07)&13.37(07)&12.53(02)&12.11(01)&11.15(11)&10.61(17)&10.07(10)\\
2005ku& 2MASX J22594265-0000478&$\cdots$&17.76(14)&16.41(09)&15.87(06)&15.66(09)&14.72(11)&14.31(10)&13.82(11)\\
2005lu& ESO 545-G038&S0&16.26(14)&15.49(04)&14.55(29)&14.28(07)&$\cdots$&13.19(04)&12.44(06)\\
2005M& NGC 2930&S0&15.20(20)&14.32(08)&14.02(14)&13.98(12)&13.45(14)&14.45(15)&12.37(12)\\
2005mc& UGC 4414&SB&16.84(37)&14.70(05)&13.47(09)&12.86(10)&$\cdots$&11.55(02)&10.80(03)\\
2005na& UGC 3634&SB&16.14(07)&13.87(04)&12.93(04)&12.53(06)&11.71(12)&11.23(15)&10.47(14)\\
2005W& NGC 691&SA&15.64(11)&12.93(09)&12.05(13)&11.76(11)&$\cdots$&9.68(02)&8.99(02)\\
2006ax& NGC 3663&SA&15.47(12)&14.06(04)&13.09(04)&12.14(04)&12.14(12)&11.38(12)&10.79(12)\\
2006bd& UGC 6609&E&17.52(32)&14.17(07)&13.34(05)&12.90(05)&11.71(12)&11.65(13)&10.98(17)\\
2006bh& NGC 7329&SB&15.56(16)&12.32(04)&11.68(06)&11.31(04)&10.88(07)&10.37(12)&9.86(07)\\
2006br& NGC 5185&S&16.04(05)&13.66(07)&12.99(05)&12.27(05)&11.59(09)&11.15(07)&10.45(12)\\
2006D& MCG -01-33-34&SB&15.64(10)&13.67(06)&13.08(05)&12.80(06)&12.09(08)&11.67(12)&10.98(11)\\
2006ef& NGC 809&S0&16.21(12)&13.96(06)&13.17(07)&12.77(06)&$\cdots$&11.41(02)&10.72(02)\\
2006ej& NGC 191A&S0&16.55(22)&14.26(08)&13.42(06)&12.95(06)&11.92(09)&11.38(06)&10.72(08)\\
2006eq& 2MASX J21283758+0113490&$\cdots$&18.82(05)&17.33(07)&16.52(07)&16.07(07)&14.54(09)&14.23(09)&13.67(09)\\
2006et& NGC 232&SB&17.21(18)&14.05(06)&13.18(09)&12.73(07)&11.52(09)&11.19(06)&10.41(07)\\
2006ev& UGC 11758&S0&17.17(13)&14.35(02)&13.53(02)&13.10(03)&11.92(08)&11.37(06)&10.71(09)\\
2006fw& SDSS J014710.33-000848.7&&19.85(23)&18.88(07)&18.28(08)&17.84(08)&$\cdots$&16.68(12)&16.04(15)\\
2006gj& UGC 2650&SA&17.27(22)&14.98(04)&14.13(03)&13.69(04)&12.48(17)&12.03(13)&11.34(17)\\
2006gt& 2MASX J00561810-0137327&$\cdots$&19.72(28)&17.10(08)&16.26(06)&15.89(05)&15.21(13)&14.80(08)&14.14(14)\\
2006hb& MCG -04-12-34&E&16.47(08)&14.06(06)&13.08(03)&12.70(03)&11.94(08)&11.84(07)&11.47(08)\\
2006hx& 2MASX J01135716+0022171&S0&18.49(09)&16.45(04)&15.70(04)&15.32(05)&14.17(07)&13.55(07)&12.91(09)\\
2006is& APMUKS(BJ) B051529.79-235009.8&$\cdots$&$\cdots$&19.25(14)&19.02(06)&18.98(07)&18.74(04)&19.73(05)&$\cdots$\\
2006kf& UGC 2829&S0&16.57(08)&13.74(06)&12.93(04)&12.55(05)&11.38(09)&10.90(09)&10.33(11)\\
2006lu& 2MASX J09151727-2536001&$\cdots$&18.30(06)&16.29(05)&15.72(05)&15.39(05)&14.48(14)&14.04(07)&13.36(13)\\
2006mr& NGC 1316&S0&11.77(12)&9.84(05)&9.17(04)&8.66(03)&8.00(04)&6.80(05)&6.22(04)\\
2006ob& UGC 1333&SA&17.11(04)&15.18(07)&14.29(07)&13.86(07)&12.59(12)&11.93(11)&11.57(22)\\
2006os& UGC 2384&S&17.04(33)&14.43(04)&13.60(04)&13.24(04)&12.20(08)&11.66(09)&11.25(09)\\
2006py& SDSS J224142.04-000812.9&$\cdots$&20.06(08)&18.23(07)&17.47(05)&17.05(04)&$\cdots$&16.16(10)&15.50(12)\\
2006X& NGC 4321&SB&12.00(05)&11.00(05)&10.00(05)&9.90(05)&10.60(37)&9.76(12)&9.10(14)\\
2007af& NGC 5584&SB&14.92(03)&13.77(07)&13.11(07)&12.72(07)&11.74(10)&11.59(12)&11.07(10)\\
2007ai& MCG -04-38-004&SA&$\cdots$&13.75(06)&12.84(02)&12.86(03)&12.05(09)&11.31(19)&10.72(11)\\
2007al& 2MASX J09591870-1928233&$\cdots$&17.95(09)&16.11(12)&15.03(05)&14.58(04)&13.59(11)&13.16(13)&12.62(07)\\
2007as& ESO 18-G18&SA&16.17(13)&13.61(13)&13.18(04)&12.86(12)&12.10(08)&11.62(09)&11.08(08)\\
2007ax& NGC 2577&S0&15.32(11)&12.79(04)&11.96(04)&11.53(04)&10.62(16)&10.01(11)&9.39(15)\\
2007ba& UGC 9798&S0&17.13(10)&14.13(07)&13.51(08)&13.06(07)&12.20(14)&11.85(15)&11.22(08)\\
2007bc& UGC 6332&SB&16.09(12)&13.64(10)&12.57(32)&12.50(05)&11.50(22)&11.23(17)&10.56(12)\\
2007bd& UGC 4455&SB&19.25(06)&14.41(06)&13.71(06)&13.35(07)&12.30(08)&11.80(08)&11.07(07)\\
2007bm& NGC 3672&SA&13.77(04)&11.84(11)&11.15(08)&10.75(06)&10.27(09)&9.73(06)&9.09(11)\\
2007ca& MCG -02-34-61&SA&18.71(24)&14.45(05)&13.96(06)&13.73(05)&12.90(16)&12.61(18)&11.84(10)\\
2007cg& ESO 508-G75&SA&17.28(09)&15.39(02)&14.65(02)&14.22(02)&13.23(07)&12.73(09)&12.06(10)\\
2007hj& NGC 7461&S0&16.41(11)&13.97(08)&13.11(05)&12.63(07)&11.57(02)&11.05(02)&10.52(02)\\
2007hx& SDSS J020627.93-005353.0&$\cdots$&19.29(08)&16.96(06)&16.29(06)&15.82(05)&15.04(09)&14.27(07)&13.71(09)\\
2007jd& SDSS J025953.65+010936.1&$\cdots$&18.75(13)&17.35(07)&16.77(08)&16.48(06)&15.61(16)&15.18(18)&14.20(16)\\
2007jg& SDSS J032950.83+000316.0&$\cdots$&19.47(10)&18.74(08)&17.46(06)&17.49(07)&16.97(13)&17.27(17)&16.44(20)\\
2007jh& CGCG 391-014&E&17.22(08)&14.83(10)&13.85(20)&13.57(08)&12.50(13)&11.82(22)&11.29(16)\\
2007le& NGC 7721&SA&13.93(18)&12.35(04)&11.72(04)&11.39(05)&11.08(02)&10.38(09)&9.52(08)\\
2007N& MCG -01-33-012&SA&17.09(08)&14.03(09)&13.30(05)&12.96(07)&11.98(10)&11.42(10)&10.89(10)\\
2007nq& UGC 595&S&17.09(08)&14.53(10)&13.57(09)&13.10(09)&12.25(11)&11.81(12)&11.06(16)\\
2007ol& 2MASX J01372378-0018422&$\cdots$&18.96(15)&17.05(12)&16.29(12)&15.84(07)&$\cdots$&14.62(13)&13.73(15)\\
2007on& NGC 1404&E&12.94(07)&10.72(11)&10.05(09)&9.59(08)&8.45(25)&8.23(25)&7.72(26)\\
2007S& UGC 5378&SA&15.43(18)&14.15(10)&13.74(08)&13.39(09)&12.51(06)&11.77(05)&11.48(08)\\
2007st& NGC 692&SB&14.62(07)&12.84(08)&12.14(13)&11.84(07)&11.06(09)&10.71(11)&9.90(20)\\
2007ux& 2MASX J10091969+1459268&$\cdots$&17.50(16)&15.76(05)&14.99(13)&14.59(16)&13.58(11)&13.09(15)&12.23(31)\\
2008ar& IC 3284&SA&17.05(16)&15.29(08)&14.64(11)&14.25(12)&13.14(24)&12.79(21)&12.33(23)\\
2008bc& KK 1524&$\cdots$&16.18(06)&14.18(07)&13.16(11)&13.02(11)&12.43(14)&12.20(13)&11.20(15)\\
2008bi& NGC 2618&SA&16.37(06)&13.31(06)&12.53(09)&12.04(08)&10.36(53)&10.90(18)&10.40(07)\\
2008bq& ESO 308-G25&Sa&17.80(04)&14.19(01)&13.46(01)&13.06(01)&12.08(09)&11.61(09)&10.87(09)\\
2008bt& NGC 3404&SB&15.43(09)&13.35(08)&12.46(08)&11.97(07)&10.76(09)&10.32(09)&9.67(09)\\
2008bz& 2MASX J12385810+1107502&$\cdots$&19.70(22)&16.55(06)&15.79(08)&15.34(09)&14.39(10)&13.93(14)&12.98(25)\\
2008C& UGC 3611&S0&16.33(14)&14.06(07)&13.31(09)&12.88(09)&11.96(12)&11.21(13)&10.58(13)\\
2008cc& ESO 107-G4&E&14.89(05)&12.79(06)&12.01(09)&11.62(08)&10.57(12)&10.15(17)&9.49(17)\\
2008cd& NGC 5038&S0&14.32(11)&12.78(03)&12.15(06)&11.81(05)&$\cdots$&10.54(02)&9.83(01)\\
2008cf& LEDA 766647&$\cdots$&19.09(07)&17.24(09)&16.64(07)&16.42(06)&16.87(20)&15.78(18)&15.37(20)\\
2008fl& NGC 6805&E&15.44(08)&13.26(02)&12.41(03)&11.97(04)&11.08(04)&10.59(08)&9.92(12)\\
2008fp& ESO 428-G14&S0&14.04(07)&12.42(06)&11.65(08)&11.05(06)&10.36(06)&9.83(07)&9.27(06)\\
2008fr& SDSS J011149.19+143826.5&$\cdots$&20.83(14)&19.44(07)&18.77(11)&18.58(10)&18.72(15)&18.07(11)&17.76(12)\\
2008fu& ESO 480-IG-021&$\cdots$&17.83(08)&15.57(09)&14.86(12)&14.49(11)&13.57(12)&13.07(11)&12.45(09)\\
2008fw& NGC 3261&SB&14.85(04)&12.66(09)&12.03(09)&11.41(10)&10.41(17)&10.03(16)&9.34(21)\\
2008gg& NGC 539&SB&16.48(06)&14.34(05)&13.53(15)&13.24(12)&12.59(09)&11.91(08)&11.51(08)\\
2008gl& UGC 881&E&16.78(18)&14.51(07)&13.68(12)&13.22(08)&12.35(15)&11.93(12)&11.19(24)\\
2008go& 2MASX J22104396-2047256&$\cdots$&18.08(10)&15.94(05)&14.98(07)&14.41(07)&13.72(14)&12.91(40)&11.68(64)\\
2008gp& MCG +00-9-74&SA&16.25(12)&14.20(09)&13.61(09)&13.26(08)&12.55(09)&12.03(11)&11.26(23)\\
2008hj& MCG -02-01-014&SB&16.76(03)&15.08(10)&14.37(05)&14.14(06)&13.84(08)&13.36(18)&12.52(11)\\
2008hu& ESO 561-G18&SA&17.77(07)&15.07(03)&14.06(03)&13.69(05)&13.06(06)&12.56(06)&11.49(07)\\
2008hv& NGC 2765&S0&14.83(01)&13.05(02)&12.28(02)&11.88(02)&10.79(07)&10.18(24)&9.70(11)\\
2008ia& ESO 125-6&S0&15.58(04)&13.17(11)&12.38(13)&12.02(13)&11.64(06)&10.86(16)&10.35(08)\\
2008O& ESO 256-G11&S0&16.10(06)&13.56(04)&12.79(03)&12.36(03)&11.19(02)&10.53(05)&9.85(05)\\
2008R& NGC 1200&S0&15.38(13)&13.04(28)&11.90(11)&11.41(10)&10.71(16)&10.23(20)&9.56(11)\\
2009aa& ESO 570-G20&Sc&19.91(14)&15.02(07)&14.42(00)&13.67(03)&13.15(09)&12.47(16)&12.04(17)\\
2009ab& UGC 2998&SB&16.90(16)&12.76(11)&12.22(13)&11.89(12)&11.25(18)&10.75(30)&10.01(44)\\
2009ad& UGC 3236&SA&16.97(10)&14.68(09)&13.98(11)&13.43(14)&12.69(08)&12.25(15)&11.61(10)\\
2009ag& ESO 492-2&SA&16.75(04)&12.02(11)&11.62(11)&11.37(12)&11.01(08)&10.52(08)&9.90(10)\\
2009al& NGC 3388&S0&16.14(03)&14.07(06)&13.35(07)&12.82(05)&11.81(07)&11.31(06)&10.64(09)\\
2009cz& NGC 2789&S0&15.01(16)&13.22(07)&12.61(07)&12.19(08)&11.56(17)&11.05(13)&10.36(15)\\
2009D& ESO 549-G031&SA&18.93(53)&14.31(14)&13.66(13)&13.32(12)&12.51(09)&12.01(11)&11.40(09)\\
2009ds& NGC 3905&SB&14.57(08)&13.37(10)&12.22(08)&11.15(12)&11.76(13)&10.99(13)&10.45(12)\\
2009F& NGC 1725&$\cdots$&15.57(12)&13.23(08)&12.37(07)&11.67(11)&9.90(20)&9.96(25)&9.37(22)\\
2009I& NGC 1080&SA&15.26(07)&13.75(08)&12.93(06)&12.23(07)&12.09(08)&11.62(06)&11.10(08)\\
2009le& ESO 478-6&SA&13.99(08)&12.63(24)&12.08(06)&11.88(08)&10.96(22)&10.59(11)&9.99(10)\\
2009P& PGC 34730&SB&16.10(21)&15.08(09)&14.68(07)&14.43(10)&13.71(16)&13.48(14)&12.73(14)\\
2009Y& NGC 5728&SB&14.21(05)&11.85(10)&11.21(10)&10.84(09)&$\cdots$&9.18(02)&8.48(02)\\
\enddata
\end{deluxetable*}
\end{longrotatetable}

\section {CSP-I Host Galaxy Properties}
We include host galaxy stellar mass of CSP-I SNe~Ia as obtained using Z-PEG. Best-fit stellar mass are listed under $M_{best}$. Lower are upper limits of stellar masses are listed under $M_{low}$ and  $M_{best}$, respectively. We also include projected distance.

\setcounter{table}{0}

\begin{longrotatetable}
\begin{deluxetable*}{ccccc}
\tablecaption{Host Galaxy Properties \label{masstable}}
%\tablewidth{500pt}
\tabletypesize{\scriptsize}
\tablehead{
\colhead{$SN  \ Name$} & \colhead{$M_{low}$} & \colhead{$M_{best}$} &
\colhead{$M_{high}$} &
\colhead{$Projected \ distance$}\\
\hline
 &&   \colhead{$log \ M_{\rm stellar}\ (M_{\odot})$}&&
\colhead{$(kpc)$} 
} 
\startdata
2004ef & 10.63 & 10.74 & 10.80&6.88\\
2004eo & 10.97 & 11.16 & 11.29&18.92\\
2004ey & 9.76 & 10.06 & 10.41&4.37\\
2004gs & 10.57 & 10.63 & 10.70&8.64\\
2004gu & 9.74 & 10.02 & 10.19&2.46\\
2005A & 11.07 & 11.19 & 11.26&2.45\\
2005ag & 10.84 & 10.90 & 10.95&13.38\\
2005al & 10.40 & 10.49 & 10.63&4.32\\
2005am & 10.68 & 10.83 & 11.26&6.67\\
2005be & 10.22 & 10.40 & 10.53&4.85\\
2005bg & 9.77 & 9.91 & 10.09&0.35\\
2005bl & 10.81 & 11.01 & 11.20&8.35\\
2005bo & 10.30 & 10.50 & 10.61&3.99\\
2005el & 10.70 & 10.84 & 11.24&13.61\\
2005eq & 10.40 & 10.56 & 10.65&17.61\\
2005hc & 10.52 & 10.60 & 10.65&8.50\\
2005hj & 9.22 & 9.32 & 9.38&3.21\\
2005iq & 10.47 & 10.54 & 10.60&13.51\\
2005ir & 9.07 & 9.23 & 9.43&0.43\\
2005kc & 10.60 & 11.06 & 11.50&3.33\\
2005ke & 10.30 & 10.43 & 10.62&5.71\\
2005ki & 10.62 & 10.80 & 10.86&27.76\\
2005ku & 9.96 & 10.04 & 10.10&0.85\\
2005lu & 9.70 & 9.86 & 9.99&2.29\\
2005M & 8.38 & 8.80 & 8.99&3.17\\
2005mc & 10.80 & 10.95 & 11.13&3.11\\
2005na & 10.90 & 10.97 & 11.05&3.96\\
2005W & 9.73 & 9.90 & 10.00&10.30\\
2006ax & 10.76 & 10.92 & 11.04&18.75\\
2006bd & 10.60 & 10.77 & 10.80&7.82\\
2006bh & 10.31 & 10.43 & 10.49&11.86\\
2006br & 10.87 & 10.97 & 11.36&3.29\\
2006D & 9.54 & 9.76 & 9.85&2.20\\
2006ef & 10.23 & 10.35 & 10.50&9.49\\
2006ej & 10.48 & 10.63 & 10.70&3.36\\
2006eq & 9.79 & 10.19 & 10.65&9.94\\
2006et & 10.70 & 10.80 & 10.88&4.71\\
2006ev & 10.65 & 10.75 & 10.91&15.22\\
2006fw & 9.26 & 9.42 & 9.61&0.71\\
2006gj & 10.36 & 10.53 & 10.60&13.19\\
2006gt & 9.87 & 9.93 & 9.94&15.49\\
2006hb & 10.04 & 10.22 & 10.38&6.19\\
2006hx & 10.12 & 10.31 & 10.44&1.91\\
2006is & 6.68 & 7.10 & 7.41&4.47\\
2006kf & 10.59 & 10.82 & 10.90&5.10\\
2006lu & 10.22 & 10.29 & 10.37&5.06\\
2006mr & 10.79 & 11.19 & 11.31&1.91\\
2006ob & 11.08 & 11.26 & 11.33&7.08\\
2006os & 10.66 & 10.81 & 10.87&8.19\\
2006py & 8.91 & 9.10 & 9.20&0.08\\
2006X & 9.06 & 9.33 & 9.51&5.09\\
2007af & 9.23 & 9.52 & 9.98&5.25\\
2007ai & 9.76 & 9.91 & 10.18&15.98\\
2007al & 9.33 & 9.47 & 9.91&1.05\\
2007as & 10.24 & 10.36 & 10.43&5.08\\
2007ax & 10.13 & 10.23 & 10.29&0.90\\
2007ba & 10.90 & 10.99 & 11.06&9.28\\
2007bc & 10.65 & 10.73 & 10.78&13.88\\
2007bd & 10.68 & 10.82 & 10.89&5.35\\
2007bm & 10.09 & 10.27 & 10.72&1.37\\
2007ca & 9.52 & 9.69 & 9.74&8.23\\
2007cg & 10.33 & 10.37 & 10.43&1.92\\
2007hj & 10.30 & 10.45 & 10.89&4.56\\
2007hx & 10.45 & 10.55 & 10.62&20.90\\
2007jd & 9.93 & 10.13 & 10.18&6.81\\
2007jg & 7.95 & 8.10 & 8.22&6.34\\
2007jh & 10.89 & 10.99 & 11.04&4.01\\
2007le & 9.71 & 9.85 & 10.16&2.40\\
2007N & 10.10 & 10.19 & 10.57&4.11\\
2007nq & 11.15 & 11.21 & 11.28&19.54\\
2007ol & 9.68 & 9.82 & 10.02&1.70\\
2007on & 10.78 & 11.04 & 11.45&9.49\\
2007S & 9.63 & 9.98 & 10.16&3.09\\
2007st & 10.85 & 10.95 & 11.40&2.46\\
2007ux & 10.23 & 10.30 & 10.34&4.23\\
2008ar & 10.11 & 10.21 & 10.28&3.25\\
2008bc & 10.00 & 10.10 & 10.21&5.90\\
2008bi & 10.27 & 10.38 & 10.48&0.93\\
2008bq & 10.74 & 10.82 & 10.91&8.59\\
2008bt & 10.70 & 10.85 & 10.91&5.06\\
2008bz & 10.44 & 10.50 & 10.56&7.98\\
2008C & 10.35 & 10.49 & 10.55&0.99\\
2008cc & 10.40 & 10.48 & 10.64&1.77\\
2008cd & 9.37 & 9.54 & 9.67&0.84\\
2008cf & 8.59 & 8.73 & 8.97&2.36\\
2008fl & 10.87 & 10.94 & 11.01&6.15\\
2008fp & 10.15 & 10.20 & 10.63&2.15\\
2008fr & 7.58 & 7.73 & 7.89&0.76\\
2008fu & 10.66 & 10.72 & 10.80&2.70\\
2008fw & 10.31 & 10.49 & 10.61&11.59\\
2008gg & 10.63 & 10.73 & 10.82&22.91\\
2008gl & 10.81 & 10.86 & 10.92&16.76\\
2008go & 10.89 & 10.94 & 11.00&17.80\\
2008gp & 10.64 & 10.73 & 10.79&11.84\\
2008hj & 10.22 & 10.40 & 10.69&15.53\\
2008hu & 10.84 & 10.99 & 11.07&39.27\\
2008hv & 10.29 & 10.55 & 10.67&9.94\\
2008ia & 10.54 & 10.77 & 11.23&2.24\\
2008O & 11.12 & 11.33 & 11.59&5.39\\
2008R & 11.15 & 11.25 & 11.31&4.00\\
2009aa & 10.10 & 10.60 & 10.80&10.06\\
2009ab & 10.20 & 10.35 & 10.78&8.30\\
2009ad & 10.50 & 10.58 & 10.64&5.44\\
2009ag & 9.99 & 10.20 & 10.60&12.79\\
2009al & 10.66 & 10.72 & 10.77&29.32\\
2009cz & 10.68 & 10.75 & 10.84&8.00\\
2009D & 10.35 & 10.43 & 10.50&20.36\\
2009ds & 10.59 & 10.77 & 10.91&4.71\\
2009F & 10.61 & 10.79 & 11.22&3.08\\
2009I & 10.53 & 10.74 & 10.85&6.65\\
2009le & 10.66 & 10.84 & 11.31&5.80\\
2009P & 9.77 & 9.87 & 9.95&1.16\\
2009Y & 10.61 & 10.76 & 10.87&5.09\\
\enddata
\end{deluxetable*}
\end{longrotatetable}

\bibliography{bibSyedUddin}{}
\bibliographystyle{aasjournal}



\end{document}